\PassOptionsToPackage{hypertexnames = false}{hyperref}
\documentclass[acmtog]{acmart}

\AtBeginDocument{%
  }

\setcopyright{rightsretained}
\acmJournal{TOG}
\acmYear{2025} \acmVolume{44} \acmNumber{6} \acmArticle{} \acmMonth{12} \acmPrice{}\acmDOI{10.1145/3763286}

\acmSubmissionID{1176}

\acmArticle{250}

\usepackage[section]{algorithm}
\usepackage[noend]{algpseudocode}
\usepackage{amsmath}
\usepackage{bm}
\usepackage{booktabs}
\usepackage{booktabs}
\usepackage{cleveref}
\usepackage{commath}
\usepackage{contour}
\usepackage[shortlabels]{enumitem}
\usepackage[acronym]{glossaries}
\usepackage{graphicx}
\usepackage{listings}
\usepackage{multirow, multicol}
\usepackage{pgfplots}
\usepackage{physics}
\usepackage{subcaption}
\usepackage{tabularx}
\usepackage{tikz}
\usepackage{tikzfill}
\usepackage{xcolor}
\usepackage{xspace}

\usepackage[nolist]{acronym}
\usepackage{amsthm, mathtools, thmtools}
\usepackage{float}
\usepackage{nicefrac}
\usepackage{pifont}
\usepackage{placeins}
\usepackage[usestackEOL]{stackengine}
\usepackage{wrapfig}
\usepackage{xparse}

\definecolor{mplblue}{HTML}{1F77B4}

\newacronym{mh}{MH}{Metropolis--Hastings}
\newacronym{mala}{MALA}{Metropolis-adjusted Langevin algorithm}
\newacronym{hmc}{H²MC}{Hamiltonian Monte Carlo}
\newacronym{mc}{MC}{Monte Carlo}
\newacronym{mcmc}{MCMC}{Markov chain Monte Carlo}
\newacronym{erpt}{ERPT}{Energy Redistribution Path Tracing}

\newacronym{bdpt}{BDPT}{bidirectional path tracing}

\newacronym{mse}{MSE}{mean squared error}
\newacronym{mrse}{MRSE}{mean relative squared error}
\newacronym{rmse}{RMSE}{relative mean squared error}
\newacronym{mape}{MAPE}{mean absolute percentage error}

\newacronym{spp}{SPP}{samples per pixel}

\glsdisablehyper

\newcommand\declaresymbol[2]{\newcommand{#1}{\TextOrMath{$#2$\xspace}{#2}}}

\declaresymbol\targetdistribution\pi
\declaresymbol\referencemeasure\lambda
\declaresymbol\targetdensity p

\declaresymbol\regenerationdistribution\mu
\declaresymbol\regenerationdensity u

\declaresymbol\integral I
\declaresymbol\integrand f
\declaresymbol\secondintegrand g
\declaresymbol\thirdintegrand h

\declaresymbol\timedomain T
\declaresymbol\timepoint t
\declaresymbol\prevtimepoint s

\declaresymbol\point x
\declaresymbol\secondpoint y
\declaresymbol\uniform u

\declaresymbol\pointset B

\declaresymbol\sequenceindex i
\declaresymbol\secondsequenceindex j
\declaresymbol\thirdsequenceindex k
\declaresymbol\lastsequenceindex n

\declaresymbol\probabilityspace\Omega
\declaresymbol\probabilitysystem{\mathcal A}
\declaresymbol\probabilitymeasure{\operatorname P}
\declaresymbol\probability\probabilitymeasure
\declaresymbol\expectation{\operatorname E}
\declaresymbol\filtration{\mathcal F}
\declaresymbol\secondfiltration{\mathcal G}

\declaresymbol\process X
\declaresymbol\secondprocess Y
\declaresymbol\thirdprocess Z

\declaresymbol\ergodicaverage S
\declaresymbol\vanillaestimator V
\declaresymbol\wasterecyclingestimator W

\declaresymbol\acceptancefunction\alpha
\declaresymbol\proposalkernel Q
\declaresymbol\proposaldensity q
\declaresymbol\transitionkernel\kappa
\declaresymbol\largestepprobability\ell
\declaresymbol\smallstepkernel\zeta

\newcommand\eqfor{\quad\text{for }}
\newcommand\eqforall{\quad\text{for all }}

\definecolor{DarkGreen}{rgb}{0, .6, 0}
\newcommand{\AlgCommentTemplate}[2]{\hfill{\fontsize{7}{6}\selectfont\textcolor{DarkGreen}{\text{#1\;#2}}}}

\newcommand{\AlgCommentLeft}[1]{\AlgCommentTemplate{$\leftarrow$}{#1}}

\algdef{SE}[SUBALG]{Indent}{EndIndent}{}{\algorithmicend\ }%
\algtext*{Indent}
\algtext*{EndIndent}

\algrenewcommand\algorithmicrequire{\textbf{Input:}}
\algrenewcommand\algorithmicensure{\textbf{Output:}}
\algnewcommand\algorithmicforeach{\textbf{for each}}
\algdef{S}[FOR]{ForEach}[1]{\algorithmicforeach\ #1\ \algorithmicdo}

\renewcommand\algorithmicdo{}

\definecolor{code-blue}{HTML}{006EB8} 
\definecolor{code-green}{HTML}{3C8031} 

\lstdefinestyle{code-style}{
    basicstyle = \ttfamily\footnotesize,
    commentstyle= \color{code-green},
    breaklines = true,
    captionpos = b,
    keywordstyle = \ttfamily\bfseries\color{code-blue},
    numbers = left,
    numberstyle = \tiny,
    xleftmargin = 12pt,
    morekeywords = {sample}
}

\newtheorem{definition}{Definition}[section]

\let\originalleft\left
\let\originalright\right
\renewcommand{\left}{\mathopen{}\mathclose\bgroup\originalleft}
\renewcommand{\right}{\aftergroup\egroup\originalright}

\makeatletter
\newcommand{\setword}[2]{%
    \phantomsection
    #1\def\@currentlabel{\unexpanded{#1}}\label{#2}%
}
\makeatother

\makeatletter
\patchcmd{\ALG@step}{\addtocounter{ALG@line}{1}}{\refstepcounter{ALG@line}}{}{}
\newcommand{\ALG@lineautorefname}{Line}
\makeatother

\makeatletter
\def\smallunderbrace#1{\mathop{\vtop{\m@th\ialign{##\crcr
   $\hfil\displaystyle{#1}\hfil$\crcr
   \noalign{\kern3\p@\nointerlineskip}%
   {\tiny\upbracefill}\crcr\noalign{\kern3\p@}}}}\limits}
\makeatother

\declaresymbol\banachspace F
\declaresymbol\banachspaceelement f

\declaresymbol\genericsemigroup T
\declaresymbol\genericgenerator A
\declaresymbol\generator\genericgenerator

\declaresymbol\measurablespace E
\declaresymbol\measurablesystem{\mathcal E}

\declaresymbol\otherpoint y
\declaresymbol\measurableset B

\declaresymbol\genericmeasure\mu

\declaresymbol{\integrandfamily}{\mathcal F}
\declaresymbol\lpexponent p

\declaresymbol\probabilitymeasurablesystem{\mathcal A}
\declaresymbol\outcome\omega

\declaresymbol\importancedistribution{\tilde\targetdistribution}
\declaresymbol\importancecorrection h
\declaresymbol\importanceweight w

\declaresymbol\discretetimedomain{\mathbb N_0}
\declaresymbol\prevprevtimepoint r
\declaresymbol\discretetimepoint n

\declaresymbol\genericprocess X

\declaresymbol\restoreprocess X
\declaresymbol\localprocess Y
\declaresymbol\killedlocalprocess Z
\declaresymbol\concatenatedprocess X

\declaresymbol\semigroup\kappa
\declaresymbol\localsemigroup\kappa
\declaresymbol\killedlocalsemigroup Q
\declaresymbol\concatenatedsemigroup Q

\declaresymbol\localgenerator L
\declaresymbol\killedlocalgenerator K
\declaresymbol\globalgenerator G
\declaresymbol\concatenatedgenerator A

\declaresymbol\closedsubspace C
\declaresymbol\core D

\declaresymbol\markovchain M
\declaresymbol\otherchain N
\declaresymbol\exponentialchain Z

\declaresymbol\dirichletform{\mathcal F}

\declaresymbol\deadstate\Delta

\declaresymbol\lifetime\tau
\declaresymbol\lifetimesum\sigma

\declaresymbol\additivefunctional A
\declaresymbol\multiplicativefunctional M

\declaresymbol\expvariable\xi

\declaresymbol\indexset I 
\declaresymbol\indexsetelement i
\declaresymbol\otherindexsetelement j
\declaresymbol\thirdindexsetelement k

\declaresymbol\tourintegral H

\renewcommand\algorithmicdo{}

\let\originalqedsymbol\qedsymbol

\declaresymbol\acceptancetime\zeta
\declaresymbol\acceptedchain\eta

\declaresymbol\otherprocess Y

\declaresymbol\genericoperator T

\declaresymbol\restorerb\lambda

\declaresymbol\kernel\kappa
\declaresymbol\subkernel Q

\declaresymbol\projection p

\declaresymbol\otherfiltration{\mathcal G}
\declaresymbol\shift\theta

\declaresymbol\integraloperator K

\declaresymbol\transferkernel T

\declaresymbol\killingrate k
\declaresymbol\killingrateprocess\Gamma
\declaresymbol\expectedlifetime c

\declaresymbol\ratekernel\alpha

\declaresymbol\combinedchain Z

\declaresymbol\holdingrate h
\declaresymbol\expparameter\eta

\declaresymbol\event A
\declaresymbol\probabilitykernel\eta

\declaresymbol\topology\eta

\declaresymbol\rejectionprobability r

\declaresymbol\drift b
\declaresymbol\diffusion\sigma
\declaresymbol\wienerprocess W

\declaresymbol\timestepdistribution T

\newcommand{\Emmett}[6] 
{
    \filldraw[fill = green] (#1) circle (.1) #4;
    \draw (#1)
    \foreach\x in {1,...,#2} {
        -- ++(rand * #3, rand * #3)
    }
    node[right] (#6) {};
    \filldraw[fill = red] (#6) circle (.1) #5;
}

\makeatletter
\newsavebox{\measure@tikzpicture}
\NewEnviron{scaletikzpicturetowidth}[1]{%
  \def\tikz@width{#1}%
  \begin{lrbox}{\measure@tikzpicture}%
  \BODY
  \end{lrbox}%
  \pgfmathparse{#1/\wd\measure@tikzpicture}%
  \BODY
}
\makeatother

\NewDocumentCommand{\underexplanation}{mm}{%
  \underset{\underset{\scriptstyle\textnormal{\makebox[0pt]{#1}}}{\downarrow}}{#2}%
}

\pgfplotsset{compat = 1.18}

\newif\compilepgf
\newif\ifdualtrack

\setcitestyle{nosort, square}

\begin{document}

\emergencystretch 3em

\title{Jump Restore Light Transport}

\author{Sascha Holl}
\authornote{Corresponding author.}
\email{sholl@mpi-inf.mpg.de}
\orcid{0009-0004-7245-7998}
\affiliation{%
  \institution{Max Planck Institute for Informatics}
  \city{Saarbrücken}
  \country{Germany}
}
\affiliation{%
  \institution{Saarland University}
  \city{Saarbrücken}
  \country{Germany}
}

\author{Gurprit Singh}
\email{gsingh@mpi-inf.mpg.de}
\orcid{0000-0003-0970-5835}
\affiliation{%
  \institution{Advanced Micro Devices, Inc. (AMD)}
  \city{Saarbrücken}
  \country{Germany}
}

\author{Hans-Peter Seidel}
\email{hpseidel@mpi-inf.mpg.de}
\orcid{0000-0002-1343-8613}
\affiliation{%
  \institution{Max Planck Institute for Informatics}
  \city{Saarbrücken}
  \country{Germany}
}

\begin{teaserfigure}

\pgfmathsetmacro\width{834}
\pgfmathsetmacro\height{469}
\pgfmathsetmacro\aspectratio{16 / 9}

\newcommand\system{dpt}
\newcommand\scene{pool}
\newcommand\firstmethod{metropolis}
\newcommand\secondmethod{metropolis_restore}
\newcommand\thirdmethod{erpt}
\newcommand\firstmethoddisplayname{Metropolis}
\newcommand\secondmethoddisplayname{Metropolis Restore (Ours)}
\newcommand\thirdmethoddisplayname{ERPT}

\pgfmathsetmacro\croplength{.12}
\pgfmathsetmacro\firstcropxmin{.5}
\pgfmathsetmacro\firstcropymin{.2}
\pgfmathsetmacro\secondcropxmin{.8}
\pgfmathsetmacro\secondcropymin{.5}
\newcommand\firstcropcolor{red}
\newcommand\secondcropcolor{blue}

\newcommand{\SplitLeftTwoImages}[2]{
    \begin{scope}
       \clip (0, 0) -- (.7 * \aspectratio, 0) -- (.3 * \aspectratio, 1) -- (0, 1) -- cycle;
        \path[fill overzoom image = results/#1.jpg] (0, 0) rectangle ( \aspectratio, 1);
    \end{scope}
   \begin{scope}
       \clip (.7 * \aspectratio, 0) -- (\aspectratio, 0) -- (\aspectratio, 1) -- (0.3 * \aspectratio, 1) -- cycle;
       \path[fill overzoom image = results/#2.jpg] (0,0) rectangle (\aspectratio, 1);
   \end{scope}
   \begin{scope}
        \draw[black, thick] (0, 0) rectangle (\aspectratio, 1);
        \draw[draw=black,thick] (.3 * \aspectratio, 1) -- (.7 * \aspectratio, 0);
   \end{scope}
}

\newcommand{\SplitRightTwoImages}[2]{
    \begin{scope}
       \clip (0, 0) -- (.3 * \aspectratio, 0) -- (.7 * \aspectratio, 1) -- (0, 1) -- cycle;
        \path[fill overzoom image = results/#1.jpg] (0, 0) rectangle ( \aspectratio, 1);
    \end{scope}
   \begin{scope}
       \clip (.3 * \aspectratio, 0) -- (\aspectratio, 0) -- (\aspectratio, 1) -- (0.7 * \aspectratio, 1) -- cycle;
       \path[fill overzoom image = results/#2.jpg] (0,0) rectangle (\aspectratio, 1);
   \end{scope}
   \begin{scope}
        \draw[black, thick] (0, 0) rectangle (\aspectratio, 1);
        \draw[draw=black,thick] (.3 * \aspectratio, 0) -- (.7 * \aspectratio, 1);
   \end{scope}
}

\newcommand{\SplitThreeImages}[3]{
    \begin{scope}
       \clip (0, 0) -- (.2 * \aspectratio, 0) -- (.3 * \aspectratio, 1) -- (0, 1) -- cycle;
        \path[fill overzoom image = results/#1.jpg] (0, 0) rectangle ( \aspectratio, 1);
    \end{scope}
    \begin{scope}
       \clip (.2 * \aspectratio, 0) -- (0.6 * \aspectratio, 0) -- (0.7 * \aspectratio, 1) -- (0.3 * \aspectratio, 1) -- cycle;
        \path[fill overzoom image = results/#2.jpg] (0, 0) rectangle ( \aspectratio, 1);
    \end{scope}
   \begin{scope}
       \clip (.6 * \aspectratio, 0) -- (\aspectratio, 0) -- (\aspectratio, 1) -- (0.7 * \aspectratio, 1) -- cycle;
       \path[fill overzoom image = results/#3.jpg] (0,0) rectangle (\aspectratio, 1);
   \end{scope}
   \begin{scope}
        \draw[black, thick] (0, 0) rectangle (\aspectratio, 1);
        \draw[draw=black,thick] (.2 * \aspectratio, 0) -- (.3 * \aspectratio, 1);
        \draw[draw=black,thick] (.6 * \aspectratio, 0) -- (.7 * \aspectratio, 1);
   \end{scope}
}

\newcommand{\CropImage}[5]{%
    \begingroup
        \pgfmathsetmacro\length{#4 * \width}
        \pgfmathsetmacro\xminval{#2 * \width}
        \pgfmathsetmacro\xmaxval{#2 * \width + \length}
        \pgfmathsetmacro\yminval{(1 - #3) * \height - \length}
        \pgfmathsetmacro\ymaxval{(1 - #3) * \height}

        \begin{scope}
            \path[fill overzoom image = results/#1.jpg] (0, 0) rectangle (1, 1);
            \clip (0, 0) rectangle (1, 1);
            \begin{scope}[xscale = 1, yscale = 1]
                \pgfmathsetmacro{\trimleft}{\xminval}
                \pgfmathsetmacro{\trimbottom}{\height - \ymaxval}  
                \pgfmathsetmacro{\trimright}{\width - \xmaxval}
                \pgfmathsetmacro{\trimtop}{\yminval}
                
                \node at (0.5,0.5) {%
                    \includegraphics[
                        width = 2.15cm, height = 2.15cm,
                        trim = {\trimleft bp} {\trimbottom bp} {\trimright bp} {\trimtop bp},
                        clip
                    ]{results/#1.jpg}%
                };
            \end{scope}
        \end{scope}

        \begin{scope}
            \draw[#5, thick] (0, 0) rectangle (1, 1);
        \end{scope}
    \endgroup
}

\def\arraystretch{.5}
\setlength\tabcolsep{1pt}
\begin{tabularx}{\linewidth}{cccccccc}
\multicolumn{4}{c}{
\begin{tikzpicture}[scale = 5]
    \SplitThreeImages{\system/\scene/_\scene_\firstmethod_30}{\system/\scene/_\scene_\secondmethod_30}{\system/\scene/_\scene_\thirdmethod_30}
    \begin{scope}
        \draw[\firstcropcolor, ultra thick]
        (\aspectratio * \firstcropxmin, \firstcropymin)
        rectangle
        (\aspectratio * \firstcropxmin + \aspectratio * \croplength, \firstcropymin + \aspectratio * \croplength);
        
        \draw[\secondcropcolor, ultra thick]
        (\aspectratio * \secondcropxmin, \secondcropymin)
        rectangle
        (\aspectratio * \secondcropxmin + \aspectratio * \croplength, \secondcropymin + \aspectratio * \croplength);
    \end{scope}
    \begin{scope}
        \contourlength{.1em}
        \node[anchor = west] at (0.025, 0.93) {\contour{black}{\textcolor{white}\firstmethoddisplayname}};
        \node[anchor = west] at (.525, 0.93) {\contour{black}{\textcolor{white}\secondmethoddisplayname}};
        \node[anchor = east] at (1.75, 0.93) {\contour{black}{\textcolor{white}\thirdmethoddisplayname}};
    \end{scope}
\end{tikzpicture}
}
&
\multicolumn{4}{c}{
\begin{tikzpicture}[scale = 5]
    \SplitThreeImages{\system/\scene/_\scene_\firstmethod_30_error}{\system/\scene/_\scene_\secondmethod_30_error}{\system/\scene/_\scene_\thirdmethod_30_error}
\end{tikzpicture}
}
\\
\begin{tikzpicture}[scale = 2.15]
    \CropImage{\system/\scene/_\scene_\firstmethod_30}\firstcropxmin\firstcropymin\croplength{\firstcropcolor}
\end{tikzpicture}
&
\begin{tikzpicture}[scale = 2.15]
    \CropImage{\system/\scene/_\scene_\secondmethod_30}\firstcropxmin\firstcropymin\croplength{\firstcropcolor}
\end{tikzpicture}
&
\begin{tikzpicture}[scale = 2.15]
    \CropImage{\system/\scene/_\scene_\thirdmethod_30}\secondcropxmin\secondcropymin\croplength{\secondcropcolor}
\end{tikzpicture}
&
\begin{tikzpicture}[scale = 2.15]
    \CropImage{\system/\scene/_\scene_\secondmethod_30}\secondcropxmin\secondcropymin\croplength{\secondcropcolor}
\end{tikzpicture}
&
\begin{tikzpicture}[scale = 2.15]
    \CropImage{\system/\scene/_\scene_\firstmethod_30_error}\firstcropxmin\firstcropymin\croplength{\firstcropcolor}
\end{tikzpicture}
&
\begin{tikzpicture}[scale = 2.15]
    \CropImage{\system/\scene/_\scene_\secondmethod_30_error}\firstcropxmin\firstcropymin\croplength{\firstcropcolor}
\end{tikzpicture}
&
\begin{tikzpicture}[scale = 2.15]
    \CropImage{\system/\scene/_\scene_\thirdmethod_30_error}\secondcropxmin\secondcropymin\croplength{\secondcropcolor}
\end{tikzpicture}
&
\begin{tikzpicture}[scale = 2.15]
    \CropImage{\system/\scene/_\scene_\secondmethod_30_error}\secondcropxmin\secondcropymin\croplength{\secondcropcolor}
\end{tikzpicture}
\\[-.75mm]
\tiny\firstmethoddisplayname  & \tiny\secondmethoddisplayname & \tiny\thirdmethoddisplayname & \tiny\secondmethoddisplayname & \tiny\firstmethoddisplayname & \tiny\secondmethoddisplayname & ´\tiny\thirdmethoddisplayname & \tiny\secondmethoddisplayname
\end{tabularx}

    \captionsetup{skip = 0pt}
    \caption{We present a novel continuous-time Markov chain Monte Carlo (MCMC) framework that \emph{adjusts} an arbitrary family of Markov processes — used solely for \emph{local exploration} — into a global process, which is invariant with respect to a given target distribution. Crucially, our approach allows for the seamless integration of \emph{any} existing MCMC sampler for local exploration. The resulting integrated algorithm consistently outperforms the original method, offering shorter running time, lower error, and reduced variance. In the figure, we depicted an equal rendering time comparison (30\,s) of Multiplexed Primary Sample Space Metropolis Light Transport (PSSMLT) \cite{hachisuka2014multiplexed} (left), its integration into our framework (middle), and Energy Redistribution Path Tracing (ERPT) \cite{cline2005erpt} (right) for the \textsc{Swimming Pool} scene provided by \citep{rioux2020delayed}.}
    \label{fig:teaser}
\end{teaserfigure}

\begin{abstract}
Markov chain Monte Carlo (MCMC) algorithms are indispensable when sampling from a complex, high-dimensional distribution by a conventional method is intractable. Even though MCMC is a powerful tool, it is also hard to control and tune in practice. Simultaneously achieving both rapid \emph{local exploration} of the state space and efficient \emph{global discovery} of the target distribution is a challenging task.  

In this work, we introduce a novel continuous-time MCMC formulation to the computer science community. Generalizing existing work from the statistics community, we propose a novel framework for \emph{adjusting} an arbitrary family of Markov processes - used for local exploration of the
state space only - to an overall process which is invariant with respect to a target~distribution.

To demonstrate the potential of our framework, we focus on a simple, but yet insightful, application in light transport simulation. As a by-product, we introduce continuous-time MCMC sampling to the computer graphics community. We show how any existing MCMC-based light transport algorithm can be seamlessly integrated into our framework. We prove empirically and theoretically that the integrated version is superior to the ordinary algorithm. In fact, our approach will convert any existing algorithm into a highly \emph{parallelizable} variant with shorter running time, smaller error and less variance.
\end{abstract}


%
%


\maketitle

\section{Introduction}\label{sec:introduction}

In light transport simulation, the computation of high-dimensional integrals is essential and is typically performed using \gls{mc} integration. Traditionally, this method involves generating independent samples within each pixel of the image space. However, a major drawback of this approach is that samples are generated regardless of their actual contribution to the final estimate. Even if a sample has no impact on the result, it is still drawn from a predefined importance distribution, without explicitly considering the value of the target density at the sampled location beforehand.

\gls{mcmc} methods offer a way to address this inefficiency. By constructing a Markov process, sample generation can be guided to better align with the target distribution, allowing for a more structured exploration of the underlying space.

\citet{veach1997thesis} introduced the \gls{mh} algorithm, arguably the most popular and widely applicable \gls{mcmc} method, to the graphics community. Building on this pioneering work, numerous \gls{mh}-based light transport algorithms have been proposed since then. In fact, every \gls{mcmc}-based light transport algorithm is actually a \gls{mh}-variant.

\paragraph*{Traditional approaches and their limitations}

In its general formulation, \gls{mh} is a recipe for constructing a Markov chain that is invariant with respect to a desired target distribution $\targetdistribution$. The key ingredient, which the user can choose (within mild constraints), is the \emph{proposal kernel} $\smallstepkernel$. Readers unfamiliar with \gls{mh} can find a detailed explanation in \autoref{sec:metropolis-hastings}. For now, it is only important to understand that \gls{mh} internally simulates a Markov chain
$\localprocess$
with transition kernel $\smallstepkernel$. This Markov chain, uniquely determined by $\smallstepkernel$, is typically chosen so as to explore the state space as rapidly and targeted as possible. Once $\smallstepkernel$
is
for our specific application, it is essentially
$\localprocess$
that we would ideally like to use for exploration.

The problem, however, is that $\smallstepkernel$, and thus
$\localprocess$,
is generally not already 
$\targetdistribution$-invariant
and therefore
is not eventually distributed according to $\targetdistribution$.
\gls{mh} can be viewed as a procedure that \emph{adjusts}
$\localprocess$
so that the resulting chain becomes
$\targetdistribution$-invariant.
This is achieved by not blindly following each proposed state transition of
$\localprocess$
, but instead accepting or rejecting them based on an \emph{acceptance probability} that depends on
$\targetdistribution$.
Details can be found in \autoref{sec:metropolis-hastings}.

\gls{mh} is surprisingly easy to implement and often performs quite well in practice. However, there are serious issues that — among other reasons — still prevent its use in production rendering today.

Modern sampling problems, whether in generative AI or rendering, are shaped by two key factors that determine the overall efficiency
: \emph{local exploration} and \emph{global discovery}. That is, we want algorithms that explore locally in a rapid and targeted way, while also ensuring global discovery of the target distribution.

\gls{mh} variants struggle with both goals. On the one hand, the \gls{mh} chain is not only 
$\targetdistribution$-invariant,
but even 
$\targetdistribution$-reversible \cite{cinlar2011probability, ethier2009markov}.
While this is useful for theoretical analysis (e.g., due to favorable spectral properties), it leads in practice to significantly reduced convergence speed \cite{Bierkens2015nonreversible}. Reversibility causes excessive backtracking — the \gls{mh} chain frequently revisits regions it has already explored. Importantly, even if the original Markov chain
$\localprocess$
induced by the proposal kernel is nonreversible, this nonreversibility is destroyed by the \gls{mh} adjustment.

In the practically most relevant Euclidean state spaces, desired (local) exploration is typically modeled via diffusion processes, as they are particularly well-suited to describe particle motion in space. To be used within \gls{mh}, these must be time-discretized in a way that renders the resulting process a Markov chain from which the proposal kernel $\smallstepkernel$ can be extracted. The most prominent methods — which we later include in our numerical study in \autoref{sec:numerical-study} — have emerged in this way.

Ensuring global discovery is even more severe. If 
$\targetdistribution$
has
separated modes, the \gls{mh} chain may get trapped in one of them, since the \gls{mh} adjustment prevents
$\localprocess$
from escaping once inside.
Even outside such worst-case settings, effective global discovery requires local exploration to be relocated to new regions of the state space after some time. One might attempt to circumvent this issue by running multiple independent \gls{mh} chains in parallel. But this approach has its own limitations. All chains could, in theory, get stuck in some mode. Moreover, \gls{mh} suffers from \emph{start-up bias}: depending on the initial state, it may take time for
$\localprocess$
to reach the target distribution.
Consequentially, early states of the \gls{mh} chain must be discarded.
As a result, even if many chains are launched in parallel, we might quickly accumulate the desired number of samples, but — in the worst case — none of them is truly representative of the target distribution.
This issue can only be resolved through an initialization phase designed to identify suitable initial states.

Instead of relying solely on multiple parallel \gls{mh} chains, the graphics community often resorts to artificial means of addressing global discovery — for instance, by replacing the "local" proposal kernel $\smallstepkernel$ with a mixture proposal kernel that includes both small-scale and large-scale moves. We discuss this further in \autoref{sec:global-discovery}.

\paragraph*{Our novel framework and its solution to traditional limitations}

To address all of these issues, we propose a framework that subsumes and significantly generalizes \gls{mh}. Our approach builds on the recently introduced \emph{Restore} algorithm by \citet{wang2021regeneration} in the statistics community. To meet the specific requirements of our domain, we significantly extend the original framework and relax its assumptions — both on the process used for local exploration and on the target distribution. In particular, it is only through this generalization that the use of this method becomes theoretically justified for light transport simulation.

In detail, we allow the use of an arbitrary family of \emph{continuous-time} Markov processes
$\localprocess^\sequenceindex$
for local exploration. Global discovery is ensured via a novel transfer mechanism that, after a duration depending on both the local target density and elapsed time, relocates the exploration process to another region of the state space. The starting point of the new local exploration may depend on the \emph{exit point} of the previous one.

The resulting overall process is invariant with respect to the desired target distribution. In this sense, our framework can be seen as an adjustment procedure that turns an arbitrary family of Markov processes into an overall process that is invariant with respect to a given target distribution.

We highlight the following advantages of our framework: \begin{enumerate}
    \item Local exploration via \emph{arbitrary} Markov processes.
    \item Potential nonreversibility of the Markov processes is \emph{preserved}, not destroyed.
    \item In \gls{mh}, a large step proposal into a high-density region (e.g., a bright region) is likely to be accepted. However, this interrupts the ongoing exploration in the previous region, leading to a bias toward oversampling the new region. In contrast, our approach never terminates local exploration based on global criteria. Instead, we end it based solely on \emph{local conditions}: the target density in the current region and the elapsed exploration time. This allows for better balance, avoids premature focus on oversampled regions, and maintains continuity in the exploration process.
    \item If the transfer between local explorations does not depend on the exit points of their respective predecessors, all local explorations can be executed in parallel without introducing startup bias into the estimator used in this context.
\end{enumerate}

\paragraph*{Outline of this work}

In \autoref{sec:related-work}, we provide an overview of related work, with a particular focus on existing \gls{mh}-based light transport algorithms and the general-purpose \gls{mh} variants originally introduced in the statistics community that they are based on.
In \autoref{sec:mcmc}, we briefly review the fundamental principles of \gls{mcmc}, for completeness.
In \autoref{sec:metropolis-hastings}, we describe the Metropolis–Hastings algorithm. Understanding its internal mechanism is crucial for the comparisons we draw in the numerical study.
In \autoref{sec:global-discovery}, we introduce the global discovery problem in detail and provide an illustrative example. We also describe the artificial workaround that is still used in current \gls{mh}-based light transport algorithms.
In \autoref{sec:restore-algorithm}, we present our generalization of the \emph{Restore} framework. We first provide an abstract formulation that holds promise for future methods, and then concretize it into a practically implementable algorithm that can be directly compared to existing techniques.
In \autoref{sec:practical-setup}, we describe the practical setup underlying our numerical study and show how any existing \gls{mcmc}-based light transport algorithm can be transformed — by integrating it into our framework — into a highly parallelizable variant with shorter runtime, lower error, and reduced variance.
In \autoref{sec:numerical-study}, we finally present our numerical study and provide empirical results that demonstrate the superiority of our framework over traditional methods.

\section{Related work}\label{sec:related-work}

\paragraph*{Seminal work}

The rendering equation~\citep{kajiya1986rendering} is typically solved with Monte Carlo estimators such as path tracing~\citep{pharr2023pbrt} and its bidirectional variants~\citep{lafortune1996rendering,veach1994bidirectional}. 
While powerful, such estimators generate samples independently of their eventual contribution, which limits their ability to resolve difficult light transport phenomena. 
This motivated the introduction of \gls{mcmc}-based light transport algorithms by \citet{veach1997thesis}, who adapted the Metropolis--Hastings (\gls{mh}) algorithm~\citep{metropolis1953equation,hastings1970monte} to the light transport setting. 
Later, \citet{kelemen2002simple} proposed to replace the path space formulation with a Euclidean \emph{primary sample space} parameterization, which greatly simplified mutation design and made \gls{mh} practical in rendering. Since then, a wide range of \gls{mh} variants have been proposed.





\paragraph*{Diffusion-based \gls{mh}}

Some of these variants are merely the ordinary \gls{mh} algorithm, but with specific choices of the proposal kernel. Beyond classical random-walk proposals, several works in the statistics community explored proposals derived from time discretizations of stochastic diffusion dynamics. 
Methods based on Langevin \citep{roberts1996mala} or Hamiltonian dynamics \citep{duane1987hmc} incorporate gradient information to better handle local exploration in anisotropic or high-dimensional distributions. 
In rendering, these ideas were applied by \citet{li2015anisotropic}, who used Hamiltonian dynamics with both first- and second-order derivatives, and by \citet{luan2020langevin}, who showed that first-order gradients suffice to generate high-quality proposals at reduced cost. 
While effective, these approaches remain computationally demanding due to their reliance on derivative information.



\paragraph*{General-purpose \gls{mh} variants}

Other methods are genuine \gls{mh} variants that have been adapted to the light transport setting. Examples include delayed rejection~\citep{mira2001delayed,rioux2020delayed}, where the rejection of proposed moves is delayed to reduce asymptotic variance  on a sweep by sweep basis; multiple-try \gls{mh}~\citep{liu2000multiple,segovia2007multiple}, which samples a set of candidates to improve acceptance rates; charted \gls{mh}~\citep{marinari1992charted,pantaleoni2017charted}, which allows switching between parametrizations; and reversible jump \gls{mh}~\citep{green1995reversible,bitterli2017reversible}, which enables transitions between different-dimensional state spaces. 
These methods extend the flexibility of \gls{mcmc} in rendering, though they inherit the reversibility and possible backtracking behavior of \gls{mh}, which may slow down convergence.



\paragraph*{Light-transport-specific \gls{mh} variants}

In addition to general-purpose \gls{mh} variants, several techniques were specifically designed for the peculiarities of light transport. 
These include approaches that fuse multiple proposal strategies~\citep{otsu2017fusing}, exploit geometric structure~\citep{otsu2018geometry}, improve path-space exploration by representing interactions along a path by halfway vectors~\citep{hanika2015improved}, or improve specular path handling by constraining mutations locally to low-dimensional manifolds~\citep{jakob2012manifold}. 



\paragraph*{\Gls{bdpt}}

\gls{bdpt} was incorporated into \gls{mh} by \citet{hachisuka2014multiplexed}.
This formulation enables the chain to adaptively choose among multiple sampling strategies through a strategy-dependent proposal kernel, while constructing estimates in a manner similar to multiple importance sampling.
Most subsequent \gls{mcmc}-based rendering methods build on this multiplexed framework as their foundation.



\paragraph*{Stratification and global exploration}

A key difficulty in \gls{mcmc} rendering is balancing local exploration with global discovery. 
\citet{gruson2020stratified} addressed this issue by spawning separate chains across multiple strata, improving robustness in difficult transport scenarios.



\paragraph*{The basic  Restore framework}

Finally, the Restore framework that we build upon was introduced in
\citet{wang2021regeneration} as a regeneration-based approach to \gls{mcmc}. 
It was later extended by \citet{mckimm2024adaptive} to include adaptive regeneration strategies, which allow for as few regenerations as possible. Our work generalizes this framework further and enables its applicability in light transport simulation.

\section{Markov chain Monte Carlo}\label{sec:mcmc}

\subsection{Basic principle}

Given a finite measure \setword{$\targetdistribution$}{inline:target-distribution}, \gls{mcmc} is a technique for estimating the integral \begin{equation}\label{eq:integral}
    \targetdistribution\integrand:=\int\integrand\dd{\targetdistribution}
\end{equation} of a $\targetdistribution$-integrable function \setword{$\integrand$}{inline:integrand}. More precisely, it is a recipe for constructing an ergodic Markov process with invariant distribution~$\targetdistribution$.

\subsubsection{Markov process}

A process is a state system evolving over time. In this work, the time domain $\timedomain$ will either be discrete, $\timedomain=\mathbb N_0$, or continuous, $\timedomain=[0,\infty)$. Informally, the process is said to be Markov, if at any fixed point in time, the evolution of the process does only depend on the present state, but not on the past. 

\subsubsection{Invariance}

$\targetdistribution$ being an invariant distribution of a Markov process $(\process_\timepoint)_{\timepoint\in\timedomain}$ is equivalent to enforcing that once $(\process_\timepoint)_{\timepoint\in\timedomain}$ is distributed according to $\targetdistribution$ at a certain time point $\prevtimepoint\in\timedomain$, every state $\process_\timepoint$ at a future time point $\timepoint\in\timedomain\cap(\prevtimepoint,\infty]$ will be distributed according to $\targetdistribution$ as well. That is, the distribution of a state is stationary in time after it once coincided with $\targetdistribution$.

\subsubsection{Ergodicity}

The ergodicity, on the other hand, will ensure that the long time average of an observation is effectively equal to space averaging with respect to the invariant distribution. That is, given that the invariant distribution $\targetdistribution$ actually exists, ergodicity is equivalent to enforcing that if $\process_0$ is distributed according to $\targetdistribution$, then \begin{align}\label{eq:ergodic-theorem}
    \ergodicaverage_\timepoint\integrand:=\frac1\timepoint\left.\begin{cases}\displaystyle\sum_{\prevtimepoint=0}^{\timepoint-1}\integrand(\process_\prevtimepoint)&\text{, if }\timedomain=\mathbb N_0\\\displaystyle\int_0^\timepoint\integrand(\process_\prevtimepoint)\dd{\prevtimepoint}&\text{, if }\timedomain=[0,\infty)\end{cases}\right\}\xrightarrow{\timepoint\to\infty}\targetdistribution\integrand
\end{align} almost surely for all $\integrand\in\mathcal L^1(\targetdistribution)$.

This characterization of ergodicity is known as \emph{Birkhoff's ergodic theorem} \cite[Theorem~25.6]{kallenberg2021probability}. In light of \eqref{eq:ergodic-theorem}, it is evident why $(\ergodicaverage_\timepoint\integrand:\timepoint\in\timedomain\setminus\{0\})$ is usually called the \emph{ergodic average estimator} of $\targetdistribution\integrand$. In this work, we will always assume that the processes under consideration exhibit this form of ergodicity. For a technical conditional ensuring \eqref{eq:ergodic-theorem} we refer to \citet{meyn1993markov}.

\subsection{Operator-theoretic viewpoint}

Invariance of Markov processes and related convergence properties towards an invariant distribution can be established much more elegantly from a functional-analytic and operator-theoretic perspective. Since in \autoref{sec:restore-algorithm} we will describe the construction of a Markov process with a prescribed invariant distribution, we adopt this perspective here and briefly introduce the central concepts needed in the remainder of this work.

Readers primarily interested in the practical aspects of the light transport algorithm may wish to skip this section on their first read and return to it later for the mathematical details.

\subsubsection{Transition semigroup}

The central object in the functional-analytic treatment of Markov processes is the \emph{transition semigroup} of the process. Formally, the transition semigroup of a Markov process $(\process_\timepoint)_{\timepoint\in\timedomain}$ with state space $\measurablespace$ is given by a \emph{regular version $\semigroup_\timepoint$ of the conditional distribution of $\process_\timepoint$ given $\process_0$} \citep[Definition~8.28]{klenke2020probability}, for all $\timepoint\in\timedomain$, such that the family $(\semigroup_\timepoint)_{\timepoint\in\timedomain}$ is an \emph{operator semigroup}. This means that $\semigroup_\timepoint(\;\cdot\;,\pointset)\circ\process_0$ is a version \citep[Definition~21.1]{klenke2020probability} of $\probability[\process_\timepoint\in\pointset\mid\process_0]$ for all measurable $\pointset\subseteq\measurablespace$, $\semigroup_\timepoint(\;\cdot\;,\pointset)$ is measurable for all measurable $\pointset\subseteq\measurablespace$, and $\semigroup_\timepoint(\point,\;\cdot\;)$ a probability measure for all $\point\in\measurablespace$.

Moreover, any measure $\targetdistribution$ can be viewed as an operator (or functional) on the space $\measurablesystem_b$ of bounded measurable $\integrand:\measurablespace\to\mathbb R$ (endowed with the supremum norm) through \eqref{eq:integral}. In complete analogy, each $\semigroup_\timepoint$ acts as an operator in this space via
\begin{equation}\label{eq:operator-semigroup}
	\semigroup_\timepoint\integrand:=\int\integrand(\secondpoint)\semigroup_\timepoint(\;\cdot\;,\dd{\secondpoint}).
\end{equation}

Without practical restrictions \citep[Corollary~11.3]{kallenberg2021probability}, it can be assumed that the operator family $(\semigroup_\timepoint)_{\timepoint\in\timedomain}$ satisfies the semigroup property
\begin{equation}\label{eq:semigroup-property}
	\semigroup_{\prevtimepoint+\timepoint}=\semigroup_\prevtimepoint\semigroup_\timepoint\eqforall\prevtimepoint,\timepoint \in \timedomain,
\end{equation}
which is known as the \emph{Chapman–Kolmogorov equation}. 

\subsubsection{Distribution and invariance}

Given the distribution of $\process_0$, the transition semigroup $(\semigroup_\timepoint)_{\timepoint\in\timedomain}$ uniquely determines the distribution of the whole Markov process $(\process_\timepoint)_{\timepoint\in\timedomain}$ \citep[Proposition~11.2]{kallenberg2021probability}, which is the reason for its central role. This observation allows the functional-analytic treatment of Markov processes by means of the extensive machinery of operator semigroup theory.

Invariance of a target distribution $\targetdistribution$ with respect to $(\process_\timepoint)_{\timepoint\in\timedomain}$ (or, more precisely, $(\semigroup_\timepoint)_{\timepoint\in\timedomain}$) is now formally defined by requiring that
\begin{equation}\label{eq:invariance-definition}
	\targetdistribution\semigroup_\timepoint=\targetdistribution\eqforall\timepoint\in\timedomain.
\end{equation}

\subsubsection{Generator}\label{sec:generator}

In general, the \emph{generator} of an operator semigroup captures its infinitesimal behavior. In our context, while the Markov process $(\process_\timepoint)_{\timepoint\in\timedomain}$ is uniquely identified by its transition semigroup $(\semigroup_\timepoint)_{\timepoint\in\timedomain}$, the key point is that — under the assumption of \emph{strong continuity} \citep[Definition~I.5.1]{engel2001semigroups} — the transition semigroup $(\semigroup_\timepoint)_{\timepoint\in\timedomain}$ is itself uniquely determined by its generator. This follows from standard generation theorems, such as the \emph{Hille–Yosida} and \emph{Lumer–Phillips} theorems \citep[Theorem~II.3.5, Theorem~II.3.15]{engel2001semigroups}.

Formally, the generator is an operator given by
\begin{equation}\label{eq:generator}
	\generator\integrand:=\begin{cases}\semigroup_1f-f&\text{, if }\timedomain=\mathbb N_0;\\\displaystyle\left.\dv{\timepoint}\semigroup_\timepoint\integrand\right|_{\timepoint=0+}&\text{, if }\timedomain=[0,\infty),\end{cases}
\end{equation}
for all $\integrand$ for which the right-hand side is well-defined. To establish invariance with respect to a target distribution $\targetdistribution$, it is practically useful to note that $\targetdistribution$ being invariant with respect to $(\process_\timepoint)_{\timepoint\in\timedomain}$ is equivalent to the condition
\begin{equation}
	\targetdistribution\generator\integrand=0
\end{equation}
for all $\integrand\in\measurablesystem_b$ in a sufficiently large \citep[Proposition~4.9.2]{ethier2009markov} class of functions. This criterion is typically much easier to verify in practice than the direct definition \eqref{eq:invariance-definition}.

\subsubsection{Adjoint operator}\label{sec:adjoint}

Let $\referencemeasure$ be a measure on $\measurablespace$ and $\genericoperator$ be an operator on $\measurablesystem_b$ with domain $\mathcal D(\genericoperator)$.  
We later also need the concept of the \emph{$\referencemeasure$-adjoint} $\genericoperator^\ast$, which is the adjoint operator of $\genericoperator$ with respect to the \emph{duality bracket} \begin{equation}
	\left\langle\integrand,\secondintegrand\right\rangle:=\int\integrand\secondintegrand\referencemeasure\eqfor\left(\integrand,\secondintegrand\right)\in\measurablesystem_b\times L^1\left(\referencemeasure\right).
\end{equation} More precisely, $\genericoperator^\ast$ is the operator on $L^1(\referencemeasure)$ with domain \begin{equation}
	\mathcal D(\genericoperator^\ast):=\left\{\secondintegrand\in L^1\left(\referencemeasure\right)\;\middle|\;\begin{split}&\exists\secondintegrand^\ast\in L^1\left(\referencemeasure\right):\\&\forall\integrand\in\mathcal D(\genericoperator):\left\langle\genericoperator\integrand,\secondintegrand\right\rangle=\bigl\langle\integrand,\secondintegrand^\ast\bigr\rangle\end{split}\right\}
\end{equation} defined by \begin{equation}
	\genericoperator^\ast\secondintegrand:=\secondintegrand^\ast.
\end{equation}

A remarkable fact is that invariance can also be characterized through the adjoint generator: If $\generator$ denotes the generator of $(\process_\timepoint)_{\timepoint\in\timedomain}$ and the target distribution $\targetdistribution$ admits a density with respect to $\referencemeasure$, then $\targetdistribution$ is invariant with respect to $(\process_\timepoint)_{\timepoint\in\timedomain}$ if and only if \begin{equation}
	\generator^\ast \secondintegrand=0
\end{equation} for all $\secondintegrand$ from a sufficiently rich subclass of $\mathcal D(\generator^\ast)$ \citep[Proposition~4.9.2]{ethier2009markov}.

\subsubsection{Lifetime restriction}\label{sec:lifetime-restriction}

In \autoref{sec:restore-algorithm}, we will describe constructions of Markov processes designed to explore the state space in a prescribed manner while remaining asymptotically distributed according to a given target distribution. To avoid trajectories that fail to converge to this distribution, it is natural to restrict their evolution to a finite time horizon rather than let them run indefinitely. Accordingly, we will consider Markov processes $(\process_\timepoint)_{\timepoint \in \timedomain}$ that are simulated only up to a finite random time~$\lifetime$, which we refer to as the \emph{lifetime} of the process.

In the continuous-time case $\timedomain=[0,\infty)$, we specifically consider lifetimes that decay with a time-dependent exponential rate
\begin{equation}
	[0,\infty)\ni\timepoint\mapsto \killingrate\left(\process_\timepoint\right),
\end{equation} where $\killingrate$ is referred to as the \emph{killing rate}. Formally, \begin{equation}
    \lifetime:=\inf\left\{\timepoint\ge0:\int_0^t\killingrate\left(\process_\prevtimepoint\right)\dd{\prevtimepoint}\ge\expvariable\right\},
\end{equation} where $\expvariable\sim\operatorname{Exp}(1)$ is independent of $(\process_\timepoint)_{\timepoint\ge0}$.

\section{Metropolis-Hastings algorithm}\label{sec:metropolis-hastings}

The
Metropolis-Hastings (MH)
algorithm is arguably the most popular and widely applicable MCMC method. It is an algorithmic construction of a Markov chain
$\markovchain$
with invariant distribution $\targetdistribution$. The procedure of simulating this chain up to a given time $\timepoint\in\mathbb N_0$ is summarized in \autoref{alg:metropolis-hastings}.

\setlength{\textfloatsep}{6pt}
\begin{algorithm}
    \caption{Metropolis-Hastings algorithm\newline with proposal kernel $\proposalkernel$ and target distribution $\targetdistribution$.}\label{alg:metropolis-hastings}
    \begin{algorithmic}[1]
        \Require{Initial state $\point_0$ and sample count $\timepoint\in\mathbb N$.}
        \Ensure{Realization $\left(\point_0,\ldots,\point_{\timepoint-1}\right)$ of the \gls{mh} chain $\markovchain$}
        \Procedure{MetropolisHastingsUpdate}{$x$}
            \State Sample $\secondpoint$ from $\proposalkernel(\point,\;\cdot\;)$;\AlgCommentLeft{generate the proposal}
            \State Sample $\uniform$ from $\mathcal U_{[0,\:1)}$;\AlgCommentLeft{uniform distribution on $[0,1)$}
            \If{$(\uniform<\acceptancefunction(\point,\secondpoint))$} \label{line:metropolis-hastings-accept}
                \State\Return$\secondpoint$;\AlgCommentLeft{with prob. $\acceptancefunction(\point,\secondpoint)$ return proposal}
            \EndIf
            \vspace{-.75mm}\State\label{line:metropolis-hastings-reject}\Return$\point$;\AlgCommentLeft{with prob. $1-\acceptancefunction(\point,\secondpoint)$ reject proposal}
        \EndProcedure
        \State\textbf{for }$\left(\prevtimepoint=1;\prevtimepoint<\timepoint;\text{++}\prevtimepoint\right)$
        \Indent
            \State\label{line:metropolis-hastings-update}$\point_\prevtimepoint\text{ = }\textproc{MetropolisHastingsUpdate}(\point_{\prevtimepoint-1})$;
        \EndIndent
    \end{algorithmic}
\end{algorithm}

\paragraph*{Algorithmic description}

The user has to specify a \setword{\emph{proposal kernel}}{inline:proposal-kernel} $\proposalkernel$. For every state $\point$, $\proposalkernel(\point,\;\cdot\;)$ is a probability measure. Now, at each discrete time step, the algorithm is \emph{proposing} a state transition candidate $\secondpoint$ drawn from $\proposalkernel(\point,\;\cdot\;)$, where $\point$ is the current state of the chain generated so far. With probability $\acceptancefunction(\point,\secondpoint)$, where $\acceptancefunction$ is an \setword{\emph{acceptance function}}{inline:acceptance-function}, the \emph{proposal} $\secondpoint$ is \emph{accepted} (line~\ref{line:metropolis-hastings-accept}) and the current state is set to $\secondpoint$. With the opposite probability, $1-\acceptancefunction(\point,\secondpoint)$, the proposal is \emph{rejected} (line~\ref{line:metropolis-hastings-reject}) and the current state will not be changed (cf. line~\ref{line:metropolis-hastings-update}).

\paragraph*{Requirements}

The initial state $\point_0$ in \autoref{alg:metropolis-hastings} may be chosen arbitrarily. The only theoretical requirement imposed on the proposal kernel $\proposalkernel$ for establishing the correctness of \autoref{alg:metropolis-hastings} is that the target distribution $\targetdistribution$ is absolutely continuous with respect to $\proposalkernel(\point,\,\cdot\,)$ for every state $\point$. This condition is intuitively reasonable, as it ensures that a proposal from $\proposalkernel(\point,\,\cdot\,)$ is able to reach any region where $\targetdistribution$ has positive measure.

\paragraph*{Acceptance function}

The mechanism ensuring that the Markov chain
$\markovchain$
is actually $\targetdistribution$-invariant is the acceptance/rejection step in \autoref{line:metropolis-hastings-accept}. The acceptance function $\acceptancefunction$ cannot be arbitrary, but there is more than one valid choice. The one usually given is optimal with respect to the \emph{Peskun-Tierney ordering} \citep{tierney1998note}.

To define it, we assume that both the target distribution $\targetdistribution$ and the proposal kernel $\proposalkernel$ admit densities with respect to a common reference measure $\referencemeasure$; that is, \begin{equation}\label{eq:targetdensity-assumption-1}
    \targetdensity_\referencemeasure:=\int\targetdensity\dd{\referencemeasure}\in(0,\infty)
\end{equation} and \begin{equation}\label{eq:target-distribution-has-density}
    \targetdistribution(\pointset)=\frac1{\targetdensity_\referencemeasure}\int_\pointset\targetdensity\dd{\referencemeasure}
\end{equation} for some density $\targetdensity$, and analogously, \begin{equation}
    \proposalkernel(\point,\pointset)=\int_\pointset\proposaldensity(\point,\;\cdot\;)\dd{\referencemeasure}
\end{equation} for some density $\proposaldensity$.
The acceptance function is then defined as \begin{equation}\label{eq:optimal-acceptance-function}
    \alpha(\point,\secondpoint):=\begin{cases}\displaystyle\min\left(1,\frac{\targetdensity(\secondpoint)\proposaldensity(\secondpoint,\point)}{\targetdensity(\point)\proposaldensity(\point,\secondpoint)}\right)&\text{, if }\targetdensity(\point)\proposaldensity(\point,\secondpoint)>0;\\1&\text{, otherwise.}\end{cases}
\end{equation}

\begin{definition}\label{def:metropolis-hastings}
    \autoref{alg:metropolis-hastings} with acceptance function \eqref{eq:optimal-acceptance-function} and the generated chain
    $\markovchain$
    are called \textbf{Metropolis-Hastings algorithm} and \textbf{Metropolis-Hastings chain with proposal kernel} $\bm\proposalkernel$ \textbf{and target distribution} $\bm\targetdistribution$, respectively.
\end{definition}

\section{Global discovery}\label{sec:global-discovery}

\emph{Local} \gls{mcmc} algorithms \textemdash\ that is, algorithms whose state transitions are confined to local neighborhoods of the current state \textemdash\ often struggle to explore multimodal target distributions. This class of algorithms includes all \gls{mh} variants whose proposal kernels propose only \emph{small-scale} moves, i.e., transitions limited to the local neighborhood of the current state.

This is a practical problem, since the most common proposal kernels for the \gls{mh} algorithm are based on time-discretized diffusion processes. They are excellent for local exploration. However, without incorporation of \emph{large-scale} moves, exploration of the whole state space will be slow and could even get stuck in local modes of the target distribution. This is particularly intuitive in the Langevin algorithm - whether Metropolis-adjusted or not - as it is effectively a stochastically perturbed gradient descent update scheme.

\paragraph*{Example}

\begin{figure}[t]
    \centering
    \begin{subfigure}{.48\linewidth}
        \centering
        \includegraphics[width = \linewidth]{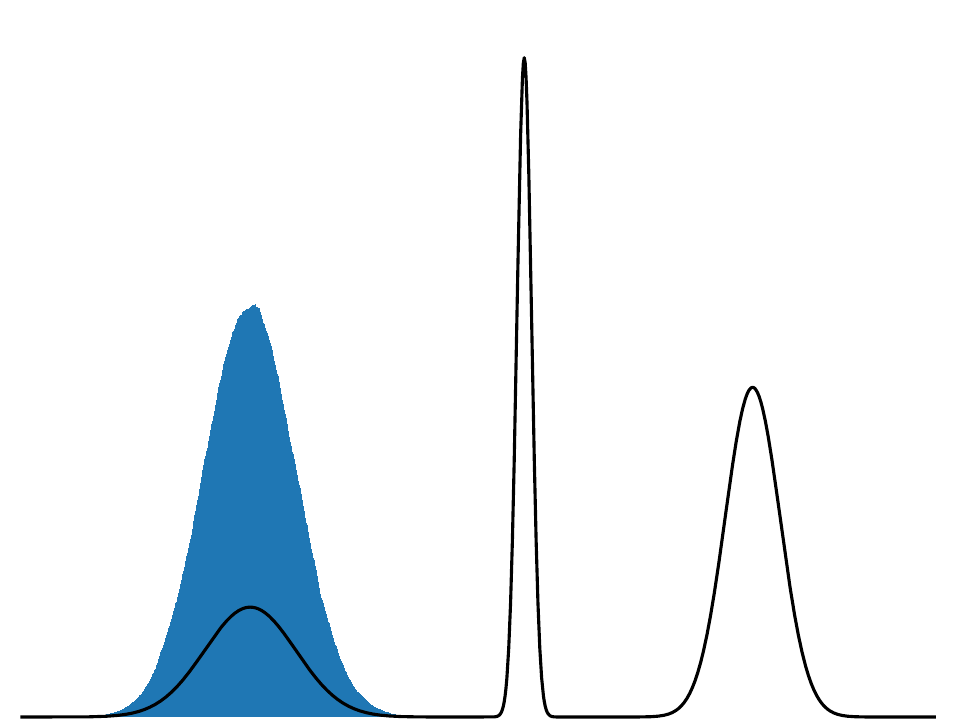}
        \caption{\footnotesize Metropolis initialized at left mode}
    \end{subfigure}\hfill
    \begin{subfigure}{.48\linewidth}
        \centering
        \includegraphics[width = \linewidth]{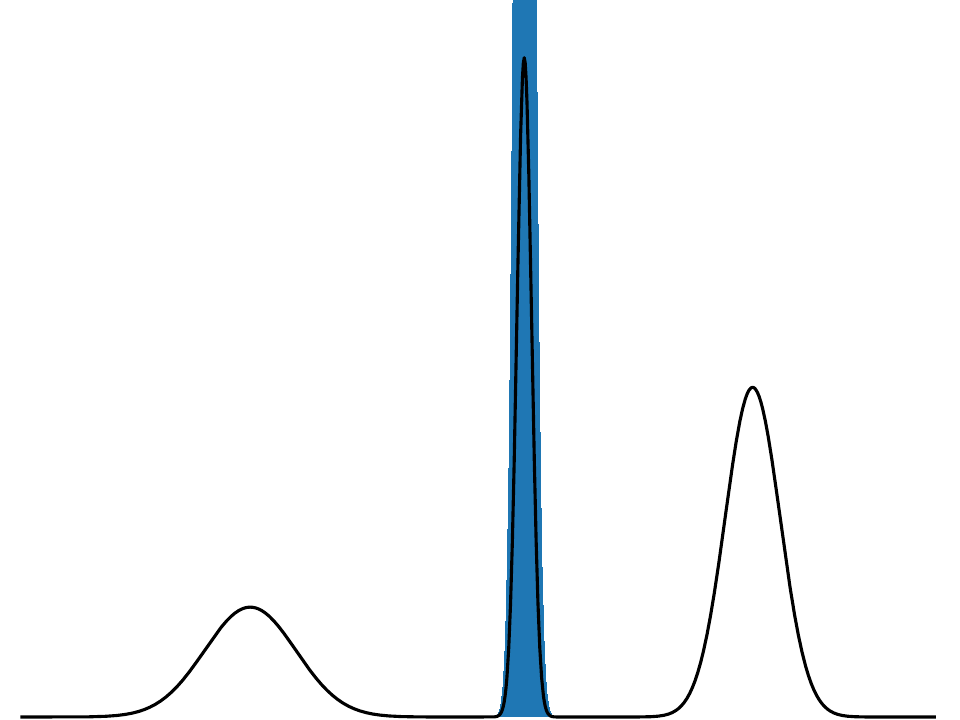}
        \caption{\footnotesize Metropolis initialized at center mode}
    \end{subfigure}

    \vspace{.5em}

    \begin{subfigure}{.48\linewidth}
        \centering
        \includegraphics[width = \linewidth]{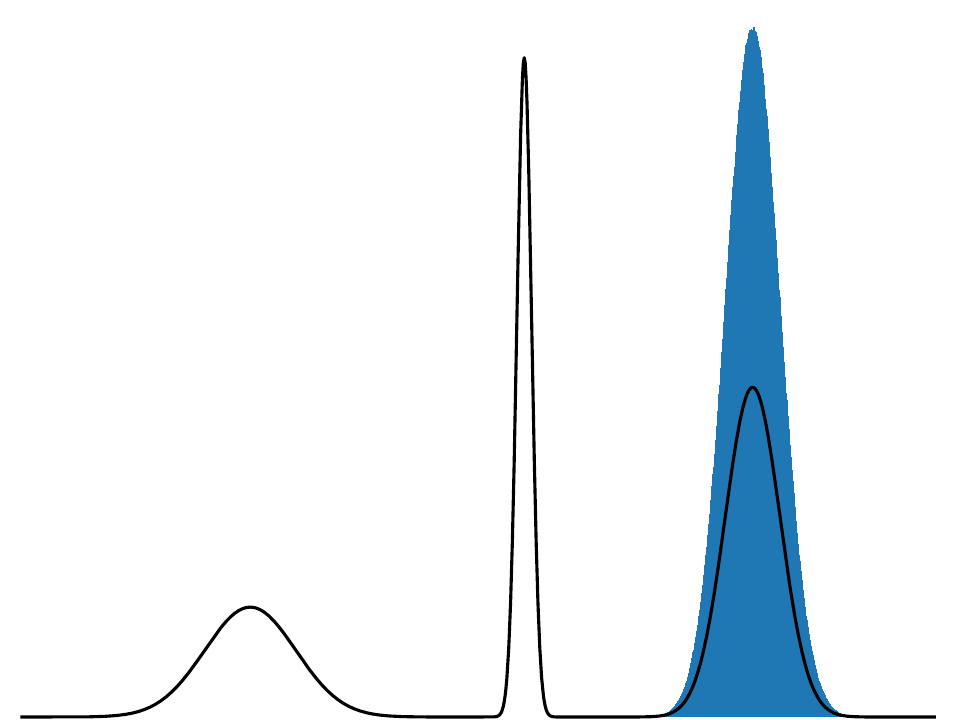}
        \caption{\footnotesize Metropolis initialized at right mode}
    \end{subfigure}\hfill
    \begin{subfigure}{.48\linewidth}
        \centering
        \includegraphics[width = \linewidth]{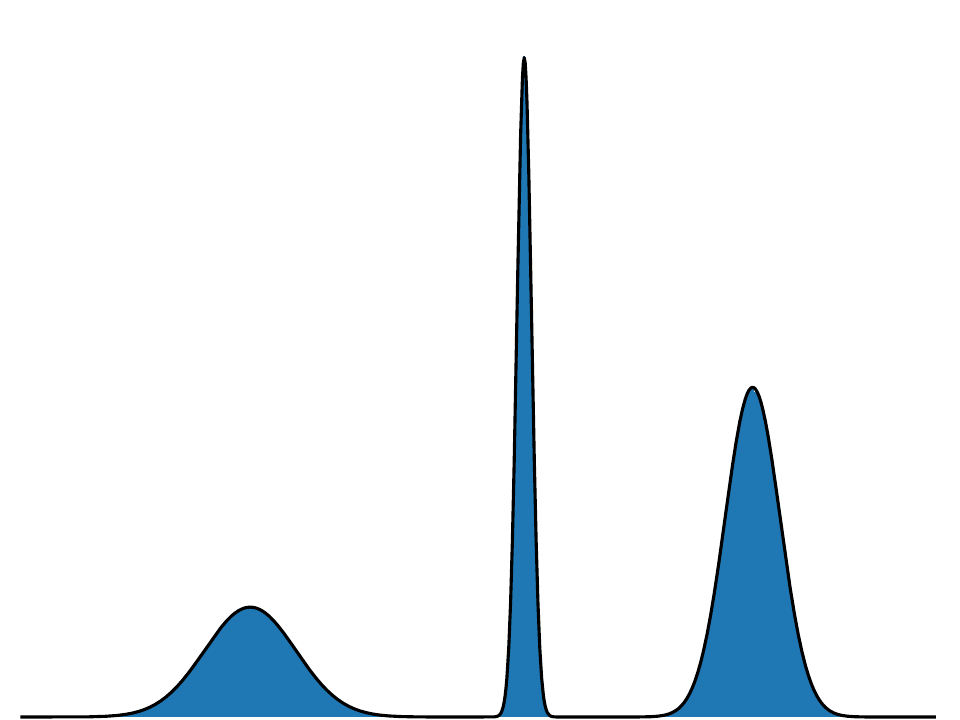}
        \caption{\footnotesize Metropolis with mixture proposal}
        \label{fig:global-discovery-d}
    \end{subfigure}

    \begin{tikzpicture}
        \begin{axis}[
            hide axis,
            xmin = 0, xmax = 1, ymin = 0, ymax = 1,
            legend columns = 2,
            legend style = {
                draw = none, fill = none,
                /tikz/every even column/.append style = {column sep = 1em}
            },
            font = \footnotesize,
        ]
        
        \addlegendimage{black, thick}
        \addlegendentry{Target density $\targetdensity$}
        
        \addlegendimage{area legend, fill = mplblue, draw = black}
        \addlegendentry{Metropolis Histogram}
        \end{axis}
    \end{tikzpicture}
  
  \captionsetup{skip = 0pt}
  \caption{Histogram of Metropolis without and with a mixture proposal. Target: 1-dimensional Gaussian mixture with three modes. (a--c) Traditional Metropolis with local Gaussian proposals $\smallstepkernel(\point,\;\cdot\;)=\mathcal N_{\varsigma^2}(\point,\;\cdot\;)$ ($\largestepprobability=0$), initialized near the left, center and right mode, respectively \textemdash\ each chain gets trapped in the mode nearest to its initial state. (d) Mixture proposal $\proposalkernel(\point,\;\cdot\;)=\largestepprobability\regenerationdistribution+(1-\largestepprobability)\smallstepkernel(\point,\;\cdot\;)$ with $\regenerationdistribution$ uniform on the support and $\largestepprobability=0.3$ enables large jumps across modes; the histogram now explores all three modes.}
  \label{fig:global-discovery}
\end{figure}

To visualize the problem, we consider a Gaussian mixture distribution \textemdash\ with three significantly separated modes \textemdash\ in \autoref{fig:global-discovery}. In (a--c), we generated $10^8$ samples using the traditional Metropolis algorithm \textemdash\ that is, the \gls{mh} algorithm with a proposal kernel $\smallstepkernel$ given by Gaussian perturbations of the current state, with a fixed variance $\varsigma^2$, of the current state. Depending on the initial state chosen, only a single mode is discovered. This does not come to a surprise. By \eqref{eq:optimal-acceptance-function}, the proposed small-scale moves are rejected when they are close to leaving the relevant support of the mode. This is simply due to the rapidly decreasing target density value.

\paragraph*{Mixture proposal}

Especially in light transport simulation, the traditional attempt to address the aforementioned issue and thereby ensure global exploration is to mix in large-scale proposals. For this reason, a typical proposal kernel $\proposalkernel$ is constructed as a \emph{mixture}: \begin{equation}\label{eq:mixture-proposal-kernel}
    \proposalkernel_\largestepprobability(\point,\;\cdot\;):=\largestepprobability\underexplanation{\textnormal{global}}\regenerationdistribution+(1-\largestepprobability)\underexplanation{\textnormal{local}}\smallstepkernel(\point,\;\cdot\;)
\end{equation} Intuitively, $\largestepprobability$, $\regenerationdistribution$, and $\smallstepkernel$ are referred to as the \emph{large step probability}, \emph{large step distribution}, and \emph{small step kernel} of the proposal scheme, respectively. By construction, $\largestepprobability \in [0,1]$ controls the probability of performing a \emph{large step} drawn from $\regenerationdistribution$. In order for
this
proposal to be effective, $\regenerationdistribution$ should be capable of making large jumps across the state space, while $\smallstepkernel$ should focus on thorough local exploration.

In our example in \autoref{fig:global-discovery-d}, we replaced the local Gaussian kernel with a mixture of the same local Gaussian kernel and a large step distribution $\regenerationdistribution$, given by a uniform distribution over the depicted domain, with a large step probability of 
$\largestepprobability=0.3$.

\paragraph*{Practical limitations}

Even though this approach works quite well in practice, it is difficult to prescribe a universally effective choice for the large step probability $\largestepprobability$. Moreover, according to \eqref{eq:mixture-proposal-kernel}, a \emph{large step proposal} introduces a large-scale move \emph{uninformatively} — that is, without taking into account how productive the current phase of exploration is. While such proposals can be rejected if the target density at the proposed state is significantly lower than at the current state, they may nonetheless divert the exploration away from regions where sampling was proceeding efficiently. Additionally, upon rejection, the chain remains at the previous state, which reduces the overall exploration speed.


\section{The Restore framework}\label{sec:restore-algorithm}

In this section, we present a generalization of the \emph{Restore} algorithm introduced by \citet{wang2021regeneration}. Our formulation extends the method by allowing for state-dependent global dynamics and by permitting each local exploration to be driven by its own Markov process. In addition, we relax the theoretical assumptions required for correctness, enabling application to the light transport setting.

\paragraph*{Core idea}

\begin{figure}[!h]
    \newcommand\radius{1.7}
    \newcommand\spacing{.7}
    
    \centering
    \scalebox{.9}{
    \begin{tikzpicture}
        \begin{scope}
            \pgfmathsetseed{5}
            \node[draw, circle, minimum size = 2*\radius cm, label = 270 : {$\localprocess^1_{\timepoint};\timepoint\in\left[0,\lifetimesum_1\right)$}] at (-\radius-\spacing/2, 0) {};
            \clip (-\radius-\spacing/2, 0) circle[radius = \radius];
            \Emmett{-\radius-\spacing/2, -\radius/5}{256}{.15}{node[anchor = east, left = 1mm] {$\localprocess^1_0$}}{node[anchor = west, right = 1mm] {\color{black}{$\localprocess^1_{\lifetime_1-}$}}}{lastnode1}
        \end{scope}

        \draw[->] (lastnode1) edge[bend left = 45, dashed] node [anchor = south] {$\regenerationdistribution$} (\radius+\spacing/2-.075, \radius/5+.075);

        \begin{scope}
            \pgfmathsetseed{6}
            \node[draw, circle, minimum size = 2*\radius cm, label = 270 : {$\localprocess^2_{\timepoint-\lifetime_1};\timepoint\in\left[\lifetime_1,\lifetime_1+\lifetime_2\right)$}] at (\radius+\spacing/2, 0) {};
            \clip (\radius+\spacing/2, 0) circle[radius = \radius];
            \Emmett{\radius+\spacing/2, \radius/5}{312}{0.1}{node[anchor = east, left = 1mm] {\color{black}{$\localprocess^2_0$}}}{node[anchor = east, left = 1mm] {\color{black}{$\localprocess^2_{\lifetime_2-}$}}}{lastnode2}
        \end{scope}
    \end{tikzpicture}
    }
    \captionsetup{skip = 0pt}
    \caption{
    We introduce a novel continuous-time MCMC framework that adjusts an arbitrary family of Markov processes $\localprocess^\sequenceindex$ \textemdash\ used solely for local exploration \textemdash\ to an overall process which is invariant with respect to a target distribution.
    Global discovery is achieved through a transfer mechanism $\regenerationdistribution$. This mechanism interrupts local exploration immediately before an exponential clock $\lifetime_\sequenceindex$ --- whose rate is inversely proportional to the target density --- expires, and then transfers the local exploration to a different region of the state space.
    }
    \label{fig:transfer}
\end{figure}

The core idea is to simulate a Markov process $\localprocess^1$ \textemdash\ which does not need to be invariant with respect to the desired target distribution \textemdash\ for a finite time $\lifetime_1$
, called \emph{lifetime} of the simulation.
A single simulation up to this lifetime is called a \emph{tour} of the process. After the simulation has been terminated, the next tour of a (possibly, but not necessarily, different) Markov process $\localprocess^2$ is spawned and simulated up to his lifetime $\lifetime_2$. The spawn location of the next tour is drawn from a user-defined distribution $\regenerationdistribution_1(\localprocess^1_{\lifetime_1-},\;\cdot\;)$, where \begin{equation}
    \localprocess^1_{\lifetime_1-}:=\lim_{\timepoint\to\lifetime_1-}\localprocess^1_\timepoint
\end{equation} denotes the \emph{exit point} \textemdash\ that is, the last point being visited immediately before termination \textemdash\ of the previous tour. The transfer rule $\regenerationdistribution_1$ can, but does not need to, depend on the exit point. By the transfer rule $\regenerationdistribution_1$, the local exploration is transferred from one portion of the space to another. This is illustrated in \autoref{fig:transfer}.

\paragraph*{The Restore process}

Continuing the idea above, the overall process is given by \begin{equation}\label{eq:concatenation}
    \concatenatedprocess_\timepoint:=\localprocess^\sequenceindex_{\timepoint-\lifetimesum_{\sequenceindex-1}}\eqfor\timepoint\in\left[\lifetimesum_{\sequenceindex-1},\lifetimesum_\sequenceindex\right),
\end{equation} where
$\lifetimesum_\lastsequenceindex:=\sum_{\sequenceindex=1}^\lastsequenceindex\lifetime_\sequenceindex$
is the time elapsed after the $\lastsequenceindex$th instance has been executed. The spawn location of the $\sequenceindex$th tour is drawn from  $\regenerationdistribution_{\sequenceindex-1}(\localprocess^{\sequenceindex-1}_{\lifetime_{\sequenceindex-1}-},\;\cdot\;)$.

\subsection{\gls{mh} as a special instance}\label{sec:metropolis-hastings-as-a-Restore-instance}

\gls{mh}
can be viewed as a special case of the above construction. To this end, we revisit the interpretation outlined in \autoref{sec:introduction}, according to which \gls{mh} essentially simulates a (conceptually \emph{local}) Markov chain $\localprocess$, implicitly defined by the proposal rule.

Let $\lifetimesum_\secondsequenceindex$ denote the time of the $\secondsequenceindex$th rejection. Then, the \gls{mh} chain naturally decomposes into tours as in \eqref{eq:concatenation}, where the transfer rules $\regenerationdistribution_\sequenceindex$ are given by a Dirac kernel: upon rejection, \gls{mh} restarts the Markov chain $\localprocess$ from its previous state.

Now consider specifically \gls{mh} with a mixture proposal rule of the form \eqref{eq:mixture-proposal-kernel}. In this case, the interpretation of local and global dynamics can be embedded directly into our framework. Indeed: Let $\lifetimesum_\secondsequenceindex$ denote the time at which \gls{mh} accepts the $\secondsequenceindex$th large step proposal. Then again, the \gls{mh} chain decomposes into tours according to \eqref{eq:concatenation}, but this time the transfer rules $\regenerationdistribution_\sequenceindex$ are all identical to the large step distribution $\regenerationdistribution$. Consequently, the tours correspond to local explorations with finite lifetime, which are \emph{killed} by the rejection step and globally \emph{revived} at a new location according to $\regenerationdistribution$.

\begin{table}
    \centering
    \small
    \caption{
        Commonly used notations throughout the paper.
    }
    \label{tab:Notation}
    \vspace{-1.5mm}
    \setlength{\tabcolsep}{2.5pt}
    \fontsize{6.9pt}{7pt}
    \begin{tabularx}{\columnwidth}{l X}
        \toprule
        \small\textbf{Notation} & \small\textbf{Description} \\
        \midrule
        \hyperref[eq:ergodic-theorem]{$\ergodicaverage$}&Ergodic average estimator\\
        $\referencemeasure$&Reference measure\\
        $\targetdistribution$&Target (probability) distribution\\
        $\targetdensity$&Density of $\targetdistribution$ with respect to $\referencemeasure$\\
        $\regenerationdistribution$&Global dynamics or large step distribution\\
        $\globalgenerator_\sequenceindex$&Generator of the global dynamics $\regenerationdistribution_\sequenceindex$ of the $\sequenceindex$th tour\\
        $\localprocess_\sequenceindex$&Local dynamics of the $\sequenceindex$th tour\\
        $\localgenerator_\sequenceindex$&Generator of the local dynamics $\localprocess^\sequenceindex$ of the $\sequenceindex$th tour\\
        $\lifetime_\sequenceindex$&Lifetime of the $\sequenceindex$th tour\\
        $\lifetimesum_\secondsequenceindex$&Sum of the lifetimes $\lifetime_1,\ldots,\lifetime_\secondsequenceindex$ of the first $\secondsequenceindex$ tours\\
        $\killingrate_\sequenceindex$&Killing rate of the $\sequenceindex$th tour\\
        \hyperref[eq:concatenation]{$\concatenatedprocess$}&Restore process\\
        $\concatenatedgenerator$&Generator of the Restore process $\concatenatedprocess$\\
        $\markovchain$&Local dynamics or \gls{mh} chain\\
        $\proposalkernel$&Proposal kernel of the \gls{mh} algorithm\\
        $\proposaldensity$&Density of $\proposalkernel$ with respect to $\referencemeasure$\\
        $\acceptancefunction$&Acceptance function of the \gls{mh} algorithm\\
        $\smallstepkernel$&Local dynamics or small step distribution
    \end{tabularx}
\end{table}

\subsection{Ensuring invariance}

The concatenation of local processes $\localprocess^\sequenceindex$ after arbitrary lifetimes $\lifetime_\sequenceindex$ is, of course, not guaranteed to be invariant with respect to a desired target distribution $\targetdistribution$.
To obtain a parameterizable control mechanism for the lifetimes $\lifetime_\sequenceindex$, we follow the general theory developed in \citet{blumenthal1968markov} and \citet{sharpe1988markov} and model the $\lifetime_\sequenceindex$ as clocks that decay according to a prescribed \emph{killing rates} $\killingrate_\sequenceindex$, as described in \autoref{sec:lifetime-restriction}.


Building on the observation of \citet{wang2021regeneration} in a simpler setting, we note that the killing rates $\killingrate_\sequenceindex$ can be chosen so that the concatenated process $\concatenatedprocess$ is invariant with respect to the desired target distribution $\targetdistribution$.
%
As in \autoref{sec:metropolis-hastings}, we assume that $\targetdistribution$ admits a density $\targetdensity$ with respect to a reference measure $\referencemeasure$. Under this assumption, the choice \begin{equation}\label{eq:killing-rate}
    \killingrate_\sequenceindex:=\frac{\left(\localgenerator_\indexsetelement^\ast+\expectedlifetime_\sequenceindex \regenerationdistribution_\indexsetelement^\ast\right)\targetdensity}{\targetdensity},
\end{equation}
ensures that the concatenated process $\concatenatedprocess$ is invariant with respect to $\targetdistribution$. A formal proof of this claim can be found in Section A of the supplemental material of this work.

In \eqref{eq:killing-rate}, the $\expectedlifetime_\sequenceindex$ are (strictly) positive constants that must be chosen such that the $\killingrate_\sequenceindex$ are (strictly) positive. Moreover, $\localgenerator_\sequenceindex$ denotes the \emph{generator} (see \autoref{sec:generator}) of the Markov process $\localprocess^\sequenceindex$ and $\localgenerator_\sequenceindex^\ast$ and $\regenerationdistribution_\sequenceindex^\ast$ denote the \emph{adjoint operator} (see \autoref{sec:adjoint}) of $\localgenerator_\sequenceindex$ and $\regenerationdistribution_\sequenceindex$ with respect to $\referencemeasure$, respectively.

\begin{definition}
    Given the choice \eqref{eq:killing-rate} of the killing rates $\killingrate_\sequenceindex$, we refer to the concatenated process $\concatenatedprocess$ as the \textbf{Restore process with local dynamics} $\bm{\localprocess^{\!\sequenceindex}}$\textbf{, global dynamics} $\bm{\regenerationdistribution_{\!\sequenceindex}}$\textbf{, and target distribution} $\bm{\targetdistribution}$.
\end{definition}

The choice \eqref{eq:killing-rate} for the killing rates $\killingrate_\sequenceindex$ is intuitively plausible, as it is inversely proportional to the target density $\targetdensity$. Accordingly, the killing rates $\killingrate_\sequenceindex$ are large in regions where the target density $\targetdensity$ is small and small where $\targetdensity$ is large.

The proof technique used in \citet{wang2021regeneration} for the special case considered there required assumptions on the local process $\localprocess$ and the target density $\targetdensity$ that render the resulting procedure unsuitable for many applications \textemdash\ including light transport simulation. In the supplementary
we provide a proof that applies in our, by far, more general setting and avoids such unnecessary restrictions.

\paragraph*{Expected lifetime}

For practical implementations \textemdash\ especially on GPUs \textemdash\ it is useful to note that the constants $\expectedlifetime_\sequenceindex$, which are the only remaining degrees of freedom in the definition \eqref{eq:killing-rate} of the killing rates $\killingrate_\sequenceindex$, are precisely the inverses of the expected lifetimes $\lifetime_\sequenceindex$ of the $\sequenceindex$th tour: \begin{equation}
    \expectation_{\regenerationdistribution_\sequenceindex}\left[\lifetime_\sequenceindex\right]=\frac1{\expectedlifetime_\sequenceindex},
\end{equation} where $\expectation_{\regenerationdistribution_\sequenceindex}$ denotes expectation with respect to the probability measure under which the local dynamics $\localprocess^\sequenceindex$ are Markov and the initial state $\localprocess^\sequenceindex_0$ is distributed according to $\regenerationdistribution_\sequenceindex$.

\paragraph*{Invariant local dynamics}

If the local processes $\localprocess^\sequenceindex$ are already invariant with respect to the target distribution $\targetdistribution$, then \begin{equation}\label{eq:invariant-local-dynamics}
    \localgenerator_\sequenceindex^\ast\targetdensity=0;
\end{equation} see \autoref{sec:adjoint}. In this case, the definition \eqref{eq:killing-rate} of the killing rates $\killingrate_\sequenceindex$ simplifies considerably — especially since in practice, the operator $\localgenerator_\sequenceindex$ is often a computationally expensive integral operator.

\paragraph*{State-independent global dynamics}

If the transfer rules $\regenerationdistribution_\sequenceindex$ do not depend on the exit point of the previous tour \textemdash\ that is, if they are simple probability distributions \textemdash\ and if they admit a density $\regenerationdensity_\sequenceindex$ with respect to the same reference measure $\referencemeasure$ under which the target distribution $\targetdistribution$ admits the density $\targetdensity$, it is straightforward to verify \begin{equation}\label{eq:state-independent-global-dynamics}
    \regenerationdistribution_\sequenceindex^\ast\secondintegrand=\regenerationdensity_\sequenceindex\int\secondintegrand\dd{\referencemeasure}.
\end{equation}

In this case, the killing rates simplify to \begin{equation}
    \killingrate_\sequenceindex=\frac{\localgenerator^\ast\targetdensity+\overbrace{\expectedlifetime_\sequenceindex\targetdensity_\referencemeasure}^{=:\;\tilde\expectedlifetime_\sequenceindex}\regenerationdensity_\sequenceindex}\targetdensity,
\end{equation} where the normalization constant $\targetdensity_\referencemeasure$ is absorbed into the user-definable constant $\expectedlifetime_\sequenceindex$, yielding the alternative user-definable constant $\tilde\expectedlifetime_\sequenceindex$. Controlling $\tilde\expectedlifetime_\sequenceindex$ instead of $\expectedlifetime_\sequenceindex$ can be advantageous for estimation, as we will see in \autoref{sec:restore-estimation} below, although this comes at the cost of losing direct control over the expected lifetimes of the tours.
If, additionally, \eqref{eq:invariant-local-dynamics} is satisfied, the killing rates further simplify to \begin{equation}\label{eq:killing-rate-simplified}
    \killingrate_\sequenceindex=\underbrace{\expectedlifetime_\sequenceindex\targetdensity_\referencemeasure}_{=\;\tilde\expectedlifetime_\sequenceindex}\frac{\regenerationdensity_\sequenceindex}{\targetdensity}.
\end{equation}
\vspace{-.5cm}

\subsection{Estimation}\label{sec:restore-estimation}

An estimator of $\targetdistribution\integrand$ for a $\targetdistribution$-integrable integrand $\integrand$ can be obtained from \eqref{eq:ergodic-theorem} as
\begin{equation}\label{eq:restore-ergodic-average}
	\frac1{\lifetimesum_\secondsequenceindex}\int_0^{\lifetimesum_\secondsequenceindex}\integrand(\concatenatedprocess_\timepoint)\dd{\timepoint}\approx\targetdistribution\integrand
\end{equation}
provided that $\secondsequenceindex$ (and thus $\lifetimesum_\secondsequenceindex$) is sufficiently large. By construction,
\begin{equation}\label{eq:tour-integral}
	\int_0^{\lifetimesum_\secondsequenceindex}\integrand(\concatenatedprocess_\timepoint)\dd{\timepoint}=\sum_{\sequenceindex=1}^\secondsequenceindex\int_0^{\lifetime_\sequenceindex}\integrand\bigl(\localprocess^\sequenceindex_\timepoint\bigr)\dd{\timepoint}.
\end{equation}

Assuming the global dynamics $\regenerationdistribution_\sequenceindex$ are state-independent and noting that the expected lifetime $\expectation_{\regenerationdistribution_\sequenceindex}\left[\lifetime_\sequenceindex\right]$ of the $\sequenceindex$th tour is given by \begin{equation}
    \expectation_{\regenerationdistribution_\sequenceindex}\left[\lifetime_\sequenceindex\right]=\frac1{\expectedlifetime_\sequenceindex}=\frac{\targetdensity_\referencemeasure}{\tilde\expectedlifetime_\sequenceindex},
\end{equation} we can estimate the normalization constant $\targetdensity_\referencemeasure$ by choosing all $\tilde\expectedlifetime_\sequenceindex$ equal and forming \begin{equation}\label{eq:normalization-constant-estimate}
    \targetdensity_\referencemeasure=\tilde\expectedlifetime_\sequenceindex\expectation_{\regenerationdistribution_\sequenceindex}\left[\lifetime_\sequenceindex\right]\approx\frac{\tilde\expectedlifetime_1}\secondsequenceindex\sum_{\sequenceindex=1}^\secondsequenceindex\lifetime_\sequenceindex=\frac{\tilde\expectedlifetime_1}\secondsequenceindex\lifetimesum_\secondsequenceindex
\end{equation} for sufficiently large $\secondsequenceindex$; cf. \citet[Section~2.2]{mckimm2024adaptive}. Combining \eqref{eq:restore-ergodic-average} with \eqref{eq:normalization-constant-estimate} yields \begin{equation}
    \frac{\tilde\expectedlifetime_1}{\secondsequenceindex\targetdensity_\referencemeasure}\int_0^{\lifetimesum_\secondsequenceindex}\integrand(\concatenatedprocess_\timepoint)\dd{\timepoint}\approx\targetdistribution\integrand
\end{equation} and hence \begin{equation}
    \frac{\tilde\expectedlifetime_1}\secondsequenceindex\int_0^{\lifetimesum_\secondsequenceindex}\integrand(\concatenatedprocess_\timepoint)\dd{\timepoint}\approx\referencemeasure(\targetdensity\integrand),
\end{equation} which is particularly useful in rendering, where the integrand typically has the form $\integrand=\secondintegrand/\targetdensity$, since the normalization constant $\targetdensity_\referencemeasure$ no longer needs to be computed \textemdash\ unlike in all previous \gls{mcmc}-based light transport algorithms, where it must be estimated in a bootstrapping phase.

\paragraph*{Practical implementation}

The main difficulties in practical implementations lie in computing the term $\localgenerator_\sequenceindex^\ast\targetdensity$ in the definition \eqref{eq:killing-rate} of the killing rate $\killingrate_\sequenceindex$, and in simulating the lifetime $\lifetime_\sequenceindex$ itself. The former can be avoided by choosing local exploration processes $\localprocess^\sequenceindex$ that are already invariant with respect to the target distribution. The latter is more challenging. In the numerical study of this work, we avoid this issue by focusing on a specific instance of the Restore process, which we describe in \autoref{sec:jump-Restore-algorithm}, and for which a simple method for simulating $\lifetime_\sequenceindex$ exists.

\subsection{Discussion}


To summarize: in the Restore framework, the user has three degrees of freedom: \begin{enumerate}
    \item The local dynamics $\localprocess^\sequenceindex$, which govern the local exploration within each tour.
    \item The global dynamics $\regenerationdistribution_\sequenceindex$, which describe the transfer between successive local explorations (or the spawn locations in the case of state-independent transfers).
    \item The inverse $\expectedlifetime_\sequenceindex$ of the expected lifetime of the $\sequenceindex$th tour.
\end{enumerate}

Compared to \citet{wang2021regeneration}, our formulation introduces several key generalizations. First, instead of using a single Markov process to govern the whole local dynamics, we allow each local exploration to follow its own Markov process. Second, we generalize the global dynamics: the starting point of the next local exploration can depend on the the exit point of the previous one. In contrast, \citet{wang2021regeneration} used a fixed distribution for this step, which justified the interpretation of the algorithm as repeatedly \emph{regenerating} a single Markov process. Finally, we relax the technical assumptions required for correctness, broadening the method’s applicability \textemdash\ in particular, to domains such as light transport. We stress that correctness of our more general formulation can still be proven rigorously as we show in Section A of the supplemental material of this work.




\section{The \emph{Jump} Restore algorithm}\label{sec:jump-Restore-algorithm}

To practically demonstrate the potential of the Restore framework presented in \autoref{sec:restore-algorithm}, this work aims to show how any existing \gls{mcmc}-based light transport algorithm can be made more efficient simply by integrating it into the Restore framework.

As described in \autoref{sec:global-discovery}, existing approaches consist of local and global components of exploration. In this work, we empirically \textemdash\ and in the supplementary, theoretically \textemdash\ demonstrate that, given \emph{any} \gls{mcmc}-based light transport algorithm, using its local component as the local dynamics and its global component as the global dynamics of the Restore process, the resulting Restore process is more efficient than the baseline algorithm in its default form with both (local and global) components.

Since every existing \gls{mcmc}-based light transport algorithm simulates a \emph{discrete-time} Markov chain, while the Restore process \textemdash\ due to its use of time-dependent exponential killing rates \textemdash\ relies on a \emph{continuous-time} formulation, we begin by describing how to embed a given discrete-time Markov chain into continuous time in such a way that its dynamics remain unaltered.

\subsection{Embedding discrete-time Markov chains into continuous time}

It is natural to embed a discrete-time chain into continuous time by \emph{holding} the states for a (random) continuous duration. And indeed, as elementary results (as found in \citet{kallenberg2021probability}) show, using exponentially distributed holding times yields a time-homogeneous and Markovian process if the original chain was. If we choose the exponential distribution parameter to be $1$, then the discrete-time Markov chain and the resulting continuous-time Markov process even share the same generator and thus follow the same dynamics.

\begin{definition}
    The process $\localprocess$ arising from the above embedding of a discrete-time chain $\markovchain$ into continuous time is referred to as the \textbf{continuous-time embedding of }$\bm\markovchain$.
\end{definition}

This yields the special case of the Restore framework introduced in \autoref{sec:restore-algorithm} that is central to our numerical study in \autoref{sec:numerical-study}:

\begin{definition}\label{def:jump-restore-process}
    Given a target distribution $\targetdistribution$ and a transfer rule $\regenerationdistribution$ on the same space, the Restore process with local dynamics $\localprocess$, global dynamics $\regenerationdistribution$, and target distribution $\targetdistribution$ is called the \textbf{Jump Restore process with local dynamics }$\bm\markovchain$\textbf{, global dynamics }$\bm\regenerationdistribution$\textbf{, and target distribution }$\bm\targetdistribution$.
\end{definition}

\subsection{Practical implementation}

The key observation is that the Jump Restore process \textemdash\ true to its name \textemdash\ is (like its local dynamics $\localprocess$) a \emph{pure-jump type} Markov process with transition rule \begin{equation}\label{eq:pure-jump-type-transition-rule}
    \frac1{1+\killingrate(\point)}\smallstepkernel(\point,\;\cdot\;)+\frac{\killingrate(\point)}{1+\killingrate(\point)}\regenerationdistribution
\end{equation} at current state $\point$. Here, $\killingrate$ denotes \emph{the} killing rate \eqref{eq:killing-rate}, which \textemdash\ due to the use of the same local and global dynamics across all tours \textemdash\ does not depend on the tour index $\sequenceindex$, and $\smallstepkernel$ denotes the transition rule of the Markov chain $\markovchain$. The constant $1$ in the numerators of \eqref{eq:pure-jump-type-transition-rule} corresponds to our choice of exponential distribution parameter for the holding times.

Given this insight, the Jump Restore process can be simulated just like any other pure-jump type Markov process. The implementation is given in \autoref{alg:jump-restore-algorithm}. The output consists of the visited states paired with their respective holding times.

\begin{algorithm}[h]
    \caption{Jump Restore algorithm with\\ local dynamics $\markovchain$, global dynamics $\regenerationdistribution$ and target distribution $\targetdistribution$}\label{alg:jump-restore-algorithm}
    \begin{algorithmic}[1]
        \Require{Initial state $\point_0$ and sample count $\discretetimepoint\in\mathbb N$.}
        \Ensure{Realization of the Jump Restore process $\concatenatedprocess$}
        \State\textbf{for }$\left(\indexsetelement=1;;\text{++}\indexsetelement\right)$
        \Indent
            \State\label{line:sample-holding-time}Sample $\timepoint_1$ from $\operatorname{Exp}\left(1\right)$;\AlgCommentLeft{holding time of the current state $\point_{\indexsetelement-1}$}
            \State\label{line:sample-killing-time}Sample $\timepoint_2$ from $\operatorname{Exp}\left(\killingrate\left(\point_{\indexsetelement-1}\right)\right)$;\AlgCommentLeft{time til next termination attempt}
            \State\textbf{if }$\left(\timepoint_1<\timepoint_2\right)$\label{line:killing}\AlgCommentLeft{next local state transition before termination attempt}
            \Indent
                \State$\Delta\timepoint_\indexsetelement=\timepoint_1$;
                \State\textbf{if }\label{line:return-if-sample-count-reached}$\left(\indexsetelement==\discretetimepoint\right)$
                \Indent
                    \State\Return $\left(\left(\Delta\timepoint_1,\point_0\right),\ldots,\left(\Delta\timepoint_\discretetimepoint,\point_{\discretetimepoint-1}\right)\right)$;
                \EndIndent
                \State\label{line:simulate-mixture-transition-embeddeding-if}Sample $\point_\indexsetelement$ from $\smallstepkernel\left(\point_{\indexsetelement-1},\;\cdot\;\right)$;\AlgCommentLeft{local state transition}
            \EndIndent
            \State\textbf{else}\AlgCommentLeft{termination before next local state transition}
            \Indent
                \State$\Delta\timepoint_\indexsetelement=\timepoint_2$;
                \State\textbf{if }\label{line:return-if-sample-count-reached2}$\left(\indexsetelement==\discretetimepoint\right)$
                \Indent
                    \State\Return $\left(\left(\Delta\timepoint_1,\point_0\right),\ldots,\left(\Delta\timepoint_\discretetimepoint,\point_{\discretetimepoint-1}\right)\right)$;
                \EndIndent
                \State\label{line:simulate-mixture-transition-embeddeding-else}Sample $\point_\indexsetelement$ from $\regenerationdistribution\left(\point_{\indexsetelement-1},\;\cdot\;\right)$;\AlgCommentLeft{global state transition}
            \EndIndent
        \EndIndent
    \end{algorithmic}
\end{algorithm}

\subsection{Example}

\begin{figure}[t]
    \centering
    \begin{subfigure}{\linewidth}
        \centering
        \includegraphics[width = .6\linewidth]{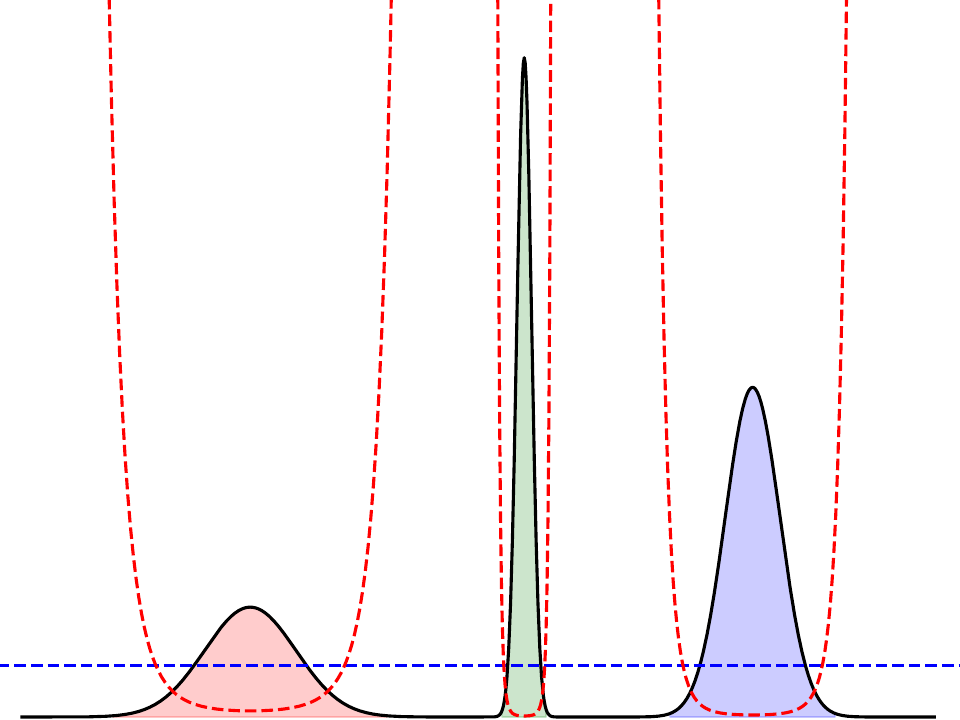}
        \ref{metropolis-restore-legend}
        \captionsetup{skip = 0pt}
        \caption{\footnotesize Multimodal target density $\targetdensity$, uniform regeneration density $\regenerationdensity$ and killing rate $\killingrate$.}
        \label{fig:metropolis-restore-a}
    \end{subfigure}\hfill

    \begin{tikzpicture}
        \begin{axis}[
            hide axis,
            xmin = 0, xmax = 1, ymin = 0, ymax = 1,
            legend columns = 2,
            legend style = {
                font = \footnotesize,
                draw = none, fill = none,
                /tikz/every even column/.append style = {column sep = 1em}
            },
            legend cell align = {left},
            legend to name = metropolis-restore-legend,
        ]
        
        \addlegendimage{black, thick}
        \addlegendentry{Target density $\targetdensity$}

        \addlegendimage{blue, thick, dashed}
        \addlegendentry{Regeneration density $\regenerationdensity$}
        
        \addlegendimage{red, thick, dashed}
        \addlegendentry{Killing rate $\killingrate$}
        \end{axis}
    \end{tikzpicture}

    \vspace{-.75em}

    \begin{subfigure}{\linewidth}
        \centering
        \includegraphics[width = \linewidth]{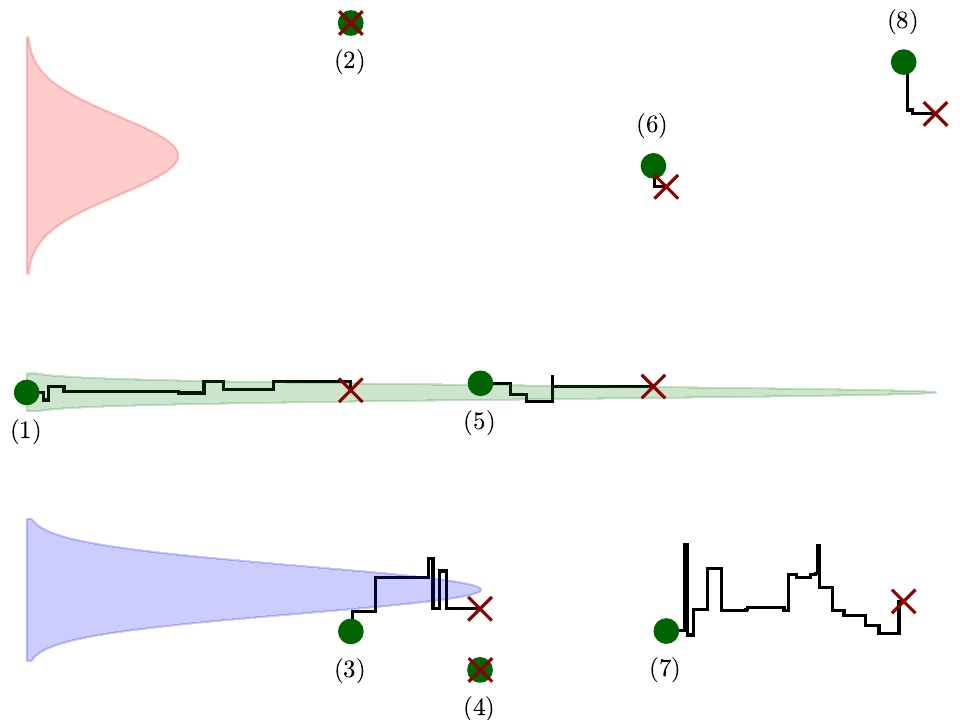}
        \captionsetup{skip = 0pt}
        \caption{\footnotesize Trajectory of the Jump Restore process, showing the first 100 state transitions. The x- and y-axes are swapped to facilitate the visualization of the exploration.}
        \label{fig:metropolis-restore-b}
    \end{subfigure}\hfill

    \captionsetup{skip = 4pt}
    \caption{Jump Restore algorithm with \gls{mh} local dynamics, uniform global dynamics and the multimodal target distribution from \autoref{fig:global-discovery}}.
    \label{fig:metropolis-restore}
\end{figure}

In \autoref{fig:metropolis-restore} we again consider the multimodal target distribution $\targetdistribution$ from \autoref{fig:global-discovery} and the \gls{mh} chain $\markovchain$ from \autoref{fig:global-discovery}(a--c). We run the Jump Restore process with local dynamics $\markovchain$, global dynamics given by the uniform distribution on the depicted domain, and target distribution $\targetdistribution$.
Since $\markovchain$ is already $\targetdistribution$-invariant, the killing rate $\killingrate$ is given by \eqref{eq:killing-rate-simplified}. The killing rate is shown in \autoref{fig:metropolis-restore-a}, together with the density $\targetdensity$ of the target distribution $\targetdistribution$ and the density $\regenerationdensity$ of the (uniform) regeneration distribution $\regenerationdistribution$.

In \autoref{fig:metropolis-restore-b} we show a single trajectory of the Jump Restore process, restricted to the first 100 state transitionss, which here constitute the first 8 tours. Tours spawned outside the relevant support of $\targetdistribution$ \textemdash\ such as tours 2 and 4 \textemdash\ are killed almost immediately, while tours like tour 1 \textemdash\ spawned at the mean of the central mode \textemdash\ have a considerably long lifetime. The constant $\expectedlifetime_\sequenceindex$ was set to $1/4$, yielding an expected tour length of $4$.

\subsection{Estimation}\label{sec:jump-restore-estimation}


Continuing from the general restore estimation methodology in \autoref{sec:restore-estimation}, and for simplicity assuming that we simulate \autoref{alg:jump-restore-algorithm} exactly until the completion of the $\secondsequenceindex$th tour \textemdash\ that is, $\sum_{\sequenceindex=1}^\secondsequenceindex\Delta\timepoint_\sequenceindex$ is a realization of $\lifetimesum_\secondsequenceindex$ \textemdash\ we obtain the practically usable approximation
\begin{equation}\label{eq:approx}
    \left(\sum_{\indexsetelement=1}^\lastsequenceindex\Delta\timepoint_\indexsetelement\right)^{-1}\sum_{\indexsetelement=1}^\lastsequenceindex\Delta\timepoint_\indexsetelement\integrand\!\left(\point_{\indexsetelement-1}\right)\approx\frac1{\lifetimesum_\secondsequenceindex}\int_0^{\lifetimesum_\secondsequenceindex}\integrand\!\left(\concatenatedprocess_\timepoint\right)\dd\timepoint\approx\int\integrand\dif\targetdistribution
\end{equation}
using the left-hand rectangle rule for integral approximation.

\section{Practical setup}\label{sec:practical-setup}

As described in \autoref{sec:jump-Restore-algorithm}, all existing \gls{mcmc}-based light transport algorithms are variants of the \gls{mh} algorithm, using a proposal kernel $\proposalkernel_\largestepprobability$ of the mixture form \eqref{eq:mixture-proposal-kernel}, which combines local and global exploration components.
We therefore assume that the reference algorithm against which we compare our method is given by an \gls{mh} chain $\markovchain^\largestepprobability$ with proposal kernel $\proposalkernel_\largestepprobability$ and target distribution $\targetdistribution$.

The key idea is to compare this reference algorithm with the Jump Restore algorithm, which reuses the local component of the reference algorithm as the local dynamics and the global component as the global dynamics in the Restore framework.
More precisely, we compare the reference algorithm $\markovchain^\largestepprobability$ with the Jump Restore process $\concatenatedprocess$, which has local dynamics $\markovchain^0$, global dynamics $\regenerationdistribution$, and target distribution $\targetdistribution$, as defined in \autoref{def:jump-restore-process}.
Here, $\regenerationdistribution$ corresponds exactly to the large step distribution used in the definition of $\proposalkernel_\largestepprobability$.

Locally, the Jump Restore process follows the same behavior as the local component of the reference algorithm, i.e., it evolves according to $\markovchain^0$. However, the length of each local exploration phase is controlled Restore's killing mechanism, which also ensures global discovery of the target distribution by triggering regeneration steps according to the global component $\regenerationdistribution$ of the reference algorithm.

\paragraph*{Convergence analysis}

In \autoref{sec:numerical-study}, we empirically compare this Jump Restore process with the Markov chain $\markovchain^\largestepprobability$. As it will turn out, ensuring global discovery by the regeneration mechanism is superior to relying on the large scale proposals affected by \eqref{eq:mixture-proposal-kernel}. We can even prove this theoretically as we do in 
Section B of the supplemental material of this work.

\section{Numerical study}\label{sec:numerical-study}

In our evaluation, we compare three representative light transport algorithms: those proposed by \citet{hachisuka2014multiplexed}, \citet{luan2020langevin}, and \citet{li2015anisotropic}. All three are ultimately grounded in classical \gls{mh} algorithms, whose proposal kernels are derived from time-discretized diffusion processes \textemdash\ Brownian motion \citep{karatzas1998brownian}, Langevin dynamics, and Hamiltonian dynamics, respectively. While these light transport methods do not implement the corresponding statistical techniques in their original form \textemdash\ for example, all of them adopt the \emph{multiplexing} technique introduced by \citet{hachisuka2014multiplexed}
\textemdash\ they still follow the principles of their general-purpose analogues.

For clarity and consistency with the 
statistical foundations, we refer to these 
light transport algorithms as \emph{Metropolis}, \gls{mala}, and \gls{hmc} 
in
this paper. Their Jump Restore variants are denoted as \emph{Metropolis Restore}, \emph{\gls{mala} Restore}, and \emph{\gls{hmc} Restore}, respectively.

\paragraph*{Parameter choices}

For all \gls{mh} baselines \textemdash\ Metropolis, \gls{mala}, and \gls{hmc}\textemdash\ we used the commonly effective choice of a large step probability $\largestepprobability=0.3$ and began estimation only after the first 10{,}000 iterations to eliminate burn-in issues.

The corresponding Jump Restore variants \textemdash\ Metropolis Restore, \gls{mala} Restore, and \gls{hmc} Restore \textemdash\ used the choice $\tilde\expectedlifetime_\sequenceindex=1$ (see \autoref{sec:restore-estimation}), which is the only remaining user-defined constant in the definition of the killing rates~\eqref{eq:killing-rate-simplified}.

\vspace{-.055cm}
\paragraph*{Rendering environment}

We implemented our method in the \texttt{pbrt} \cite{pharr2023pbrt} and \texttt{lmc} \cite{luan2020langevin} rendering system and applied the \gls{mcmc} methods to both direct and indirect lighting.

\vspace{-.055cm}
\paragraph*{Test scenes}

We evaluated
divserse scenes exhibiting different light transport characteristics. The scenes are
from three sources: \textsc{Contemporary Bathroom}, \textsc{Glass of Water}, and \textsc{Country Kitchen} from \citet{bitterli2016resources}; \textsc{Veach, Ajar} and \textsc{Torus} from \citet{lmc}; and \textsc{Swimming Pool} from \citet{rioux2020delayed}.

\vspace{-.055cm}
\paragraph*{Error metrics}

We assessed several quantitative metrics:
the $L^1$-error, $L^2$-error (i.e., \gls{mse}), \gls{mrse}, \gls{rmse}, and \gls{mape}. Reference images were generated using \gls{bdpt} with $2^{20}$ \gls{spp}. In addition, we computed the (empirical) variance of the resulting renderings.

Reference images were generated using \gls{bdpt} with $2^{20}$ \gls{spp}. In addition, we computed the
variance of the resulting renderings.

\vspace{-.055cm}
\paragraph*{Evaluation}

Qualitative comparisons are shown in \cref{fig:teaser,fig:veach,fig:torus,fig:bathroom,fig:glass}, covering all test scenes introduced above \textemdash\ except the \textsc{Country Kitchen} scene, which can be found in Figure 1 of the supplementary material of this work. They reveal that Restore variants consistently achieve better mode coverage and visual fidelity than their standard counterparts. Quantitative results in \cref{fig:veach-torus-error,fig:bathroom-glass-error} confirm this observation: both the $L^2$-error and empirical variance decrease more rapidly over time for the Restore variants.
Detailed tables with absolute error values at equal time and equal \gls{spp} are provided in Section C of the supplementary material of this work.

Restore's improvement over the \gls{mh} baselines is more pronounced in equal time than in equal \gls{spp} comparisons. Restore kills local explorations (almost) immediately when the target density (almost) vanishes, assigning those samples negligible $\Delta\timepoint_\sequenceindex$ in the estimator~\eqref{eq:approx}. While these samples contribute nothing to the estimate, they are still counted towards \gls{spp}, which causes an underestimation of Restore's efficiency in equal \gls{spp} comparisons, whereas equal time more accurately reflects practical performance.

\vspace{-.055cm}
\paragraph*{Hardware setup}

All renderings were performed on a system equipped with two AMD EPYC~7702 processors, each providing 64~cores and 128~hardware threads. The CPUs operate at a clock speed of 2–3.3\,GHz and are paired with 2048\,GiB of DDR4 ECC memory running at 3200\,MHz. All computations were executed entirely on the CPU. Since our hardware limited us to at most 256 concurrent threads, the results reported in \autoref{sec:numerical-study} may further improve on systems that support a higher degree of parallelism.

\subsection{\gls{erpt}}

\gls{erpt} \cite{cline2005erpt}
shares
conceptual similarities with the Jump Restore algorithm proposed in this work. Both methods perform sampling through multiple, short
chains that are initialized in different regions of the
space and explore their local neighborhood.

More preciely, \gls{erpt} begins with a bootstrapping phase in which a
specifiable
number of random paths
are generated and evaluated using traditional \gls{mc} integration. Each path’s contribution to the image — thought of as its "energy” in the \gls{erpt} context — is computed in this step. A subset of these paths is then selected as initial states for Markov chains, with selection biased toward higher-energy paths. These Markov chains then evolve by mutating the current path and "redistributing" its energy to the resulting path. The average number of started chains per pixel and a common fixed length for them are user-defined parameters.

Due to this conceptual similarity, we included \gls{erpt} in our numerical evaluation. 
From a theoretical perspective, however, \gls{erpt} exhibits some of the same challenges as traditional \gls{mh} sampling: it may get trapped in local modes of the target distribution, and the choice of 
chain length and count is highly scene-dependent and difficult to tune universally. Additionally, since \gls{erpt} operates on a per-pixel basis, it is
less closely related to our proposed
algorithm than \gls{mh}
. 
We explicitly refer again to our discussion in \autoref{sec:metropolis-hastings-as-a-Restore-instance}, which further clarifies how \gls{mh}
relates to
our algorithm.

\section{Conclusion}\label{sec:conclusion}

We introduced a generalized framework for \gls{mcmc} sampling that overcomes key limitations of \gls{mh}-based light transport algorithms. By decoupling local exploration and global discovery into separate mechanisms, our approach preserves desirable properties such as nonreversibility and rapid local exploration, while ensuring theoretical correctness and practical flexibility. This makes it possible to construct sampling schemes that explore efficiently, exploit local geometry or structure, and remain globally consistent — thereby bridging a longstanding gap between the needs of modern rendering applications and the limitations of classical \gls{mcmc} methods. Beyond rendering, our framework holds strong potential for a broad spectrum of industrial applications, including generative AI, where the global discovery of multimodal distributions is also fundamental.

\bibliographystyle{ACM-Reference-Format}
\bibliography{bibliography}

\newpage
\begin{figure*}
    \input{figures/veach}
    \captionsetup{skip = 0pt}
    \caption{Equal rendering time comparison (20s) of \gls{mala} (left), \gls{mala} Restore (middle), and \gls{erpt} (right) for the \textsc{Veach, ajar} scene provided by \citep{lmc}.}
    \label{fig:veach}
\end{figure*}

\begin{figure*}[t]
    \input{figures/veach_torus_error}
    \captionsetup{skip = 0pt}
    \caption{$L^2$-error and empirical variance over rendering time in seconds for the \textsc{Veach, ajar} and \textsc{Torus} scene depicted in \autoref{fig:veach} and \autoref{fig:torus}, respectively.}
    \label{fig:veach-torus-error}
\end{figure*}

\begin{figure*}
    \input{figures/torus}
    \captionsetup{skip = 0pt}
    \caption{Equal rendering time comparison (120s) of \gls{hmc} (left), \gls{hmc} Restore (middle), and \gls{erpt} (right) for the \textsc{Torus} scene provided by \citep{lmc}.}
    \label{fig:torus}
\end{figure*}

\clearpage
\begin{figure*}
    \input{figures/bathroom}
    \captionsetup{skip = 0pt}
    \caption{Equal rendering time comparison (120s) of Metropolis (left), Metropolis Restore (middle), and \gls{erpt} (right) for the \textsc{Contemporary Bathroom} scene provided by \citep{bitterli2016resources}.}
    \label{fig:bathroom}
\end{figure*}

\begin{figure*}[t]
        \centering
    \begin{subfigure}[t]{.2\textwidth}
        \centering
        \begin{tikzpicture}[scale = .48]
            \begin{axis}[
                title = {\textsc{Contemporary Bathroom}},
                title style = {yshift = -1.5ex},
                grid = major,
                legend cell align = {left},
                legend pos = north east,
                legend style = {font = \scriptsize},
                xlabel = {Rendering time in seconds},
                ylabel = {$L^2$-error},
                ymode = log,
                xmin = 0,
                xmax = 300,
                log ticks with fixed point,
                xtick = {100, 200, 300},
                xticklabels = {$100$, $200$, $300$},
                ytick = {.1, .5, 1, 2, 5, 10},
                yticklabels = {$0.1$, $0.5$, $1$, $2$, $5$, $10$},
                yminorticks = false,
                tick align = outside,
            ]
                \addplot[smooth, color = black, dashed, thick, no markers] table[x index = 1, y index = 3, col sep = tab] {results/pbrt4/bathroom/bathroom_metropolis.txt};
                \addplot[smooth, color = black, thick, no markers] table[x index = 1, y index = 3, col sep = tab] {results/pbrt4/bathroom/bathroom_metropolis_restore.txt};
                \addplot[smooth, color = magenta, dashed, thick, no markers] table[x index = 1, y index = 3, col sep = tab] {results/pbrt4/bathroom/bathroom_erpt.txt};

            \end{axis}
        \end{tikzpicture}
    \end{subfigure}
    \hspace{.5cm}
    \begin{subfigure}[t]{.2\textwidth}
        \centering
        \begin{tikzpicture}[scale = .48]
            \begin{axis}[
                title = {\textsc{Contemporary Bathroom}},
                title style = {yshift = -1.5ex},
                grid = major,
                xlabel = {Rendering time in seconds},
                ylabel = {Empirical variance},
                ymode = log,
                xmin = 0,
                xmax = 300,
                log ticks with fixed point,
                xtick = {100, 200, 300},
                xticklabels = {$100$, $200$, $300$},
                ytick = {.1, .5, 1, 2, 5, 10},
                yticklabels = {$0.1$, $0.5$, $1$, $2$, $5$, $10$},
                yminorticks = false,
                tick align = outside,
            ]
                \addplot[smooth, color = black, dashed, thick, no markers] table[x index = 1, y index = 7, col sep = tab] {results/pbrt4/bathroom/bathroom_metropolis.txt};
                \addplot[smooth, color = black, thick, no markers] table[x index = 1, y index = 7, col sep = tab] {results/pbrt4/bathroom/bathroom_metropolis_restore.txt};
                \addplot[smooth, color = magenta, dashed, thick, no markers] table[x index = 1, y index = 7, col sep = tab] {results/pbrt4/bathroom/bathroom_erpt.txt};
            \end{axis}
        \end{tikzpicture}
    \end{subfigure}
    \hspace{.5cm}
    \begin{subfigure}[t]{.2\textwidth}
        \centering
        \begin{tikzpicture}[scale = .48]
            \begin{axis}[
                title = {\textsc{Glass of Water}},
                title style = {yshift = -1.5ex},
                grid = major,
                legend cell align = {left},
                legend pos = north east,
                legend style = {font = \scriptsize},
                xlabel = {Rendering time in seconds},
                ylabel = {$L^2$-error},
                ymode = log,
                xmin = 0,
                xmax = 300,
                log ticks with fixed point,
                xtick = {100, 200, 300},
                xticklabels = {$100$, $200$, $300$},
                ytick = {.001, .01, .1, .5, 1, 2, 5},
                yticklabels = {$.001$, $0.01$, $0.1$, $0.5$, $1$, $2$, $5$},
                yminorticks = false,
                tick align = outside,
            ]
                \addplot[smooth, color = black, dashed, thick, no markers] table[x index = 1, y index = 3, col sep = tab] {results/pbrt4/glass/glass_metropolis.txt};
                \addplot[smooth, color = black, thick, no markers] table[x index = 1, y index = 3, col sep = tab] {results/pbrt4/glass/glass_metropolis_restore.txt};
                \addplot[smooth, color = magenta, dashed, thick, no markers] table[x index = 1, y index = 3, col sep = tab] {results/pbrt4/glass/glass_erpt.txt};

            \end{axis}
        \end{tikzpicture}
    \end{subfigure}
    \hspace{.5cm}
    \begin{subfigure}[t]{.2\textwidth}
        \centering
        \begin{tikzpicture}[scale = .48]
            \begin{axis}[
                title = {\textsc{Glass of Water}},
                title style = {yshift = -1.5ex},
                grid = major,
                xlabel = {Rendering time in seconds},
                ylabel = {Empirical variance},
                ymode = log,
                xmin = 0,
                xmax = 300,
                log ticks with fixed point,
                xtick = {100, 200, 300},
                xticklabels = {$100$, $200$, $300$},
                ytick = {.001, .01, .1, .5, 1, 2, 5},
                yticklabels = {$.001$, $0.01$, $0.1$, $0.5$, $1$, $2$, $5$},
                yminorticks = false,
                tick align = outside,
            ]
                \addplot[smooth, color = black, dashed, thick, no markers] table[x index = 1, y index = 7, col sep = tab] {results/pbrt4/glass/glass_metropolis.txt};
                \addplot[smooth, color = black, thick, no markers] table[x index = 1, y index = 7, col sep = tab] {results/pbrt4/glass/glass_metropolis_restore.txt};
                \addplot[smooth, color = magenta, dashed, thick, no markers] table[x index = 1, y index = 7, col sep = tab] {results/pbrt4/glass/glass_erpt.txt};
            \end{axis}
        \end{tikzpicture}
    \end{subfigure}
    \begin{tikzpicture}
        \begin{axis}[
            hide axis,
            xmin = 0,
            xmax = 1,
            ymin = 0,
            ymax = 1,
            legend columns=4,
            legend style = {draw = none, font = \footnotesize, /tikz/every even column/.append style = {column sep = .5cm}},
        ]
            \addlegendimage{black, dashed, thick}
            \addlegendentry{Metropolis}
            \addlegendimage{black, thick}
            \addlegendentry{Metropolis Restore (Ours)}
            \addlegendimage{magenta, dashed, thick}
            \addlegendentry{\gls{erpt}}
        \end{axis}
    \end{tikzpicture}
    
    \captionsetup{skip = 0pt}
    \caption{$L^2$-error and empirical variance over rendering time in seconds for the \textsc{Veach, ajar} and \textsc{Torus} scene depicted in \autoref{fig:veach} and \autoref{fig:torus}, respectively.}
    \label{fig:bathroom-glass-error}
\end{figure*}

\begin{figure*}
    \input{figures/glass}
    \captionsetup{skip = 0pt}
    \caption{Equal rendering time comparison (60s) of Metropolis (left), Metropolis Restore (middle), and \gls{erpt} (right) for the \textsc{Glass of Water} scene provided by \citep{bitterli2016resources}.}
    \label{fig:glass}
\end{figure*}

\clearpage

\ifdualtrack
\appendix

\section{Invariance of the Restore process}\label{sec:invariance-proof}

The process $\concatenatedprocess$ constructed in \autoref{sec:restore-algorithm} is \emph{locally} (during a single tour) a \emph{jump-type Markov process}. In general, such a process is a Markov process $(\concatenatedprocess_\timepoint)_{\timepoint\ge0}$ with values in a measurable space $(\measurablespace,\measurablesystem)$ and a generator (see \autoref{sec:generator}) of the form \begin{equation}
    \concatenatedgenerator\integrand:=\localgenerator\integrand+\killingrate\globalgenerator\integrand\;\;\;\text{for }\integrand\in\mathcal D(\localgenerator).
\end{equation} where $\localgenerator$ is the generator of a Markov semigroup $(\localsemigroup_\timepoint)_{\timepoint\ge0}$ on $(\measurablespace,\measurablesystem)$, $\killingrate:\measurablespace\to[0,\infty)$ is $\measurablesystem$-measurable and $\regenerationdistribution$ is a Markov kernel on $(\measurablespace,\measurablesystem)$. $\killingrate$, $\regenerationdistribution$ and \begin{equation}
    \ratekernel(\point,\measurableset):=\killingrate(\point)\regenerationdistribution(\point,\measurableset)\;\;\;\text{for }(\point,\measurableset)\in\measurablespace\times\measurablesystem
\end{equation} are called \emph{rate function}, \emph{jump transition kernel} and \emph{rate kernel} of $(\concatenatedprocess_\timepoint)_{\timepoint\ge0}$, respectively.

The terminology is intuitive: A transition according to $\ratekernel$ is happening with rate $\killingrate$ and is performed by following $\regenerationdistribution$. For a general discussion on such processes, we refer to \citet[Chapter~13]{kallenberg2021probability}.

Given a probability measure $\targetdistribution$ on $(\measurablespace,\measurablesystem)$, our goal in this section is to derive a choice for the rate function $\killingrate$ for which the process $(\concatenatedprocess_\timepoint)_{\timepoint\ge0}$ is $\targetdistribution$-invariant. The result obtained this way will immediately give the right choice of killing rates for the Restore process \eqref{eq:concatenation}.

The following assumption can be relaxed, but it simplifies the analysis and is already in place in our setting in \autoref{sec:restore-algorithm}. We assume, for the remainder of this section, that both, the regeneration kernel $\regenerationdistribution$ and the target distribution $\targetdistribution$, have a density with respect to a common $\sigma$-finite reference measure $\referencemeasure$ on $(\measurablespace,\measurablesystem)$. To be precise, we assume that there is a $\measurablesystem$-measurable $\targetdensity:\measurablespace\to(0,\infty)$ with \begin{equation}
    \targetdensity_\referencemeasure:=\referencemeasure\targetdensity\in(0,\infty)
\end{equation} and \begin{equation}
    \targetdistribution=\frac\targetdensity{\targetdensity_\referencemeasure}\referencemeasure.
\end{equation} Analogously, we assume that there is an $\measurablesystem^{\otimes2}$-measurable $\regenerationdensity:\measurablespace^2\to[0,\infty)$ with \begin{equation}
    \regenerationdensity_\referencemeasure(\point):=\referencemeasure\regenerationdensity(\point,\;\cdot\;)\in(0,\infty)
\end{equation} and \begin{equation}
    \regenerationdistribution(\point,\;\cdot\;)=\frac{\regenerationdensity(\point,\;\cdot\;)}{\regenerationdensity_\referencemeasure(\point)}\referencemeasure.
\end{equation} for all $\point\in\measurablespace$.

If $\genericoperator$ is an operator on $\measurablesystem_b$, we denote the $\referencemeasure$-adjoint of $\genericoperator$, defined in $\autoref{sec:adjoint}$, by $\genericoperator^\ast$. We are now able to state and proof the main theorem of this section:

\begin{theorem}\label{thm:restore-process-invariance}
    \normalfont If $\targetdensity\in\mathcal D(\localgenerator^\ast)$ with \begin{equation}\label{eq:restore-process-invariance-eq1}
        \referencemeasure\left(\regenerationdistribution^\ast\localgenerator^\ast\targetdensity\right)=0
    \end{equation} and \begin{equation}\label{eq:restore-process-invariance-eq2}
        \killingrate=\frac{\left(\expectedlifetime_1\regenerationdistribution^\ast+\localgenerator^\ast\right)\targetdensity}\targetdensity
    \end{equation}
    for some $\expectedlifetime_1\ge0$, then \begin{equation}
        \targetdistribution\concatenatedgenerator\integrand=0\;\;\;\text{for all }\integrand\in\mathcal D(\concatenatedgenerator).
    \end{equation}
    \begin{proof}[Proof\textup:\nopunct]
            Let $\integrand\in\mathcal D(\localgenerator)$. By \eqref{eq:restore-process-invariance-eq1}, \begin{equation}
        \begin{split}
            \left|\left\langle\regenerationdistribution\integrand,\localgenerator^\ast\targetdensity\right\rangle_\referencemeasure\right|&=\left|\left\langle\integrand,\regenerationdistribution^\ast\localgenerator^\ast\targetdensity\right\rangle_\referencemeasure\right|\\&\le\left\|\integrand\right\|_\infty\underbrace{\referencemeasure\left(\regenerationdistribution^\ast\localgenerator^\ast\targetdensity\right)}_{=\:0}=0
        \end{split}
    \end{equation} and hence \begin{equation}
        \referencemeasure\left(\targetdensity\ratekernel\integrand\right)=\underbrace{\left\langle\regenerationdistribution\integrand,\localgenerator^\ast\targetdensity\right\rangle_\referencemeasure}_{=\:0}+\:\expectedlifetime_1\left\langle\regenerationdistribution\integrand,\regenerationdistribution^\ast\targetdensity\right\rangle_\referencemeasure
    \end{equation} by \eqref{eq:restore-process-invariance-eq2}. Again by \eqref{eq:restore-process-invariance-eq2}, we obtain \begin{equation}
        \begin{split}
            \referencemeasure\left(\targetdensity\killingrate\integrand\right)&=\left\langle\integrand,\left(\localgenerator^\ast+\expectedlifetime_1\regenerationdistribution^\ast\right)\targetdensity\right\rangle_\referencemeasure\\&=\left\langle\left(\localgenerator+\expectedlifetime_1\regenerationdistribution\right)\integrand,\targetdensity\right\rangle_\referencemeasure
        \end{split}
    \end{equation} and hence \begin{equation}
        \begin{split}
            \targetdensity_\referencemeasure\targetdistribution\localgenerator\integrand&=\referencemeasure\left(\targetdensity\localgenerator\integrand\right)-\referencemeasure\left(\targetdensity\killingrate\integrand\right)+\referencemeasure\left(\targetdensity\ratekernel\integrand\right)\\&=\expectedlifetime_1\left\langle\regenerationdistribution\integrand,\tilde\localgenerator^\ast\targetdensity\right\rangle_\referencemeasure=0.
        \end{split}
    \end{equation} For the last equation, simply note that \begin{equation}
        \globalgenerator^\ast=\regenerationdistribution^\ast-\operatorname{id}_{\measurablesystem_b}
    \end{equation} and \begin{equation}
        \globalgenerator1=0.
    \end{equation}
    \end{proof}
\end{theorem}

A crucial condition, which is implicit in \eqref{eq:restore-process-invariance-eq2}, is that $\expectedlifetime_1$ must be large enough to ensure $\killingrate\ge0$. If $(\localsemigroup_\timepoint)_{\timepoint\ge0}$ is already $\targetdistribution$-invariant, then $\localgenerator^\ast\targetdensity=0$ (see \autoref{sec:adjoint}) and hence condition \eqref{eq:restore-process-invariance-eq2} is reduced to \begin{equation}
    \killingrate=\expectedlifetime_1\frac{\regenerationdistribution^\ast\targetdensity}\targetdensity
\end{equation} In that case, $\expectedlifetime_1$ can be arbitrary. If $\targetdistribution$ is state-independent (i.e. does not depend on the first argument and hence is simply a measure), then the adjoint of $\targetdistribution$ is given by \begin{equation}
    \regenerationdistribution^\ast\secondintegrand=\frac{\referencemeasure\secondintegrand}{\regenerationdensity_\referencemeasure}\regenerationdensity\eqforall\secondintegrand\in\mathcal L^1(\referencemeasure)
\end{equation} and hence condition \eqref{eq:restore-process-invariance-eq2} actually is \begin{equation}
    \killingrate=\frac{\displaystyle\expectedlifetime_1\frac{\targetdensity_\referencemeasure}{\regenerationdensity_\referencemeasure}\regenerationdensity+\localgenerator^\ast\targetdensity}\targetdensity.
\end{equation}

\section{Theoretical convergence investigation}\label{sec:convergence-analysis}

Our goal in this section is to compare the quality of an estimation based on \begin{enumerate}
    \item the \gls{mh} chain $\markovchain^\largestepprobability$ with proposal kernel $\proposalkernel_\largestepprobability$ and target distribution $\targetdistribution$; and
    \item the process $\concatenatedprocess$ formed by the jump Restore algorithm with local dynamics $\markovchain^0$ and regeneration distribution $\regenerationdistribution$.
\end{enumerate} The generators $\localgenerator_\largestepprobability$ and $\concatenatedgenerator$ of $\markovchain^\largestepprobability$ and $\concatenatedprocess$ will play an important role in the theoretical investigation of this comparison.


Our measure of quality will be based on the \emph{asymptotic variance}. If $\integrand\in L^2(\targetdistribution)$, we can define the asymptotic variance \begin{equation}
    \sigma^2_\concatenatedgenerator(\integrand):=\lim_{\timepoint\to\infty}\operatorname{Var}_\targetdistribution\left[\frac1{\sqrt\timepoint}\int_0^\timepoint\integrand(\concatenatedprocess_\timepoint)\dif\timepoint\right]
\end{equation} of the ergodic average estimation of $\targetdistribution\integrand$ based on $\concatenatedprocess$, whenever the limit exists. In that case, the ergodic average \eqref{eq:ergodic-theorem} is asymptotically behaving as $\sigma^2_\genericgenerator(\integrand)/\timepoint$. Here, the variance is taken with respect to the probability measure with respect to which the initial distribution of $\concatenatedprocess$ is already $\targetdistribution$. That is, the assumption in this consideration is that the process is started in stationarity.

In the $L^2(\targetdistribution)$-reversible case, the $L^2$\emph{-relaxation time} \citep[Section~3]{eberle2024relaxation} of a time-homogeneous Markov process $\concatenatedprocess$ with invariant measure $\targetdistribution$ is governed by the \emph{spectral gap} \begin{equation}
    \operatorname{gap}(\genericgenerator):=\inf\left\{\frac{\dirichletform(\integrand)}{\left\|\integrand\right\|_{L^2(\targetdistribution)}^2}:\integrand\in\mathcal D(\genericgenerator)\setminus\{0\}\text{ with }\targetdistribution\integrand=0\right\}.
\end{equation} of its $L^2(\targetdistribution)$-generator $\genericgenerator$. Here, \begin{equation}
    \dirichletform(\integrand,\secondintegrand):=-\langle\genericgenerator\integrand,\secondintegrand\rangle_{L^2(\targetdistribution)}\;\;\;\text{for }\integrand\in\mathcal D(\genericgenerator)\text{ and }\secondintegrand\in L^2(\targetdistribution)
\end{equation} is the \emph{Dirichlet form} associated with $\genericgenerator$ and \begin{equation}
    \dirichletform(\integrand):=\dirichletform(\integrand,\integrand)\eqfor\integrand\in\mathcal D(\genericgenerator).
\end{equation} In the general, non-$L^2(\targetdistribution)$-reversible, case, the $L^2$-relaxation time is at least still lower bounded by $\operatorname{gap}(\genericgenerator)$.

However, similar to the celebrated \emph{Peskun-Tierney ordering}, we can show \citep{andrieu2019ordering} that an ordering \begin{equation}\label{eq:dirichlet-form-ordering}
    \dirichletform_\largestepprobability(\integrand)\le\dirichletform(\integrand)\eqforall\integrand\in\measurablesystem_b
\end{equation} on the Dirichlet forms $\dirichletform_\largestepprobability$ and $\dirichletform$ of $\concatenatedprocess$ and $\markovchain^\largestepprobability$, respectively, directly corresponds to an ordering \begin{equation}
    \sigma^2_{\localgenerator_\largestepprobability}(\integrand)\ge\sigma^2_\concatenatedgenerator(\integrand)
\end{equation} of the asymptotic variances of the corresponding ergodic averages.

We still can compare the performance of different processes by considering the asymptotic variances of their ergodic averages. Similar to the celebrated \emph{Peskun-Tierney ordering} \citep{tierney1998note}, an ordering among asymptotic variances is directly related to an ordering of the corresponding \emph{Dirichlet forms} \citep{andrieu2019ordering}. 

That is, we can compare the asymptotic variances of $\markovchain^\largestepprobability$ and $\concatenatedprocess$ by proving that their corresponding Dirichlet forms $\dirichletform_\largestepprobability$ and $\dirichletform$ are ordered. We are actually able to do so:

\begin{theorem}
    \normalfont If \begin{equation}
        \expectedlifetime_1\ge\frac\largestepprobability{2\targetdensity_\referencemeasure},
    \end{equation} then \eqref{eq:dirichlet-form-ordering} is satisfied. 
    \begin{proof}[Proof\textup:\nopunct]
        First of all, it is useful to note that \begin{equation}
            \proposalkernel_\largestepprobability=\largestepprobability\proposalkernel_1+(1-\largestepprobability)\proposalkernel_0.
        \end{equation}
        Now, the transition kernel of the Metropolis-Hastings chain $(\markovchain^\largestepprobability_\discretetimepoint)_{\discretetimepoint\in\discretetimedomain}$ is given by \begin{equation}
            \kernel_\largestepprobability(\point,\measurableset):=\int_\measurableset\proposalkernel_\largestepprobability(\point,\dd{\otherpoint})\acceptancefunction_\largestepprobability(\point,\otherpoint)=\largestepprobability\kernel_1+(1-\largestepprobability)\kernel_0
        \end{equation} for $(\point,\measurableset)\in\measurablespace\times\measurablesystem$, where $\acceptancefunction_\largestepprobability$ is the acceptance function \eqref{eq:optimal-acceptance-function} of the Metropolis-Hastings algorithm with proposal kernel $\proposalkernel_\largestepprobability$ and target density $\targetdistribution$ (which effectively does not depend on $\largestepprobability$, whenever the small step kernel $\smallstepkernel$ has a \emph{symmetric} density with respect to the reference measure $\referencemeasure$) and \begin{equation}
            \rejectionprobability_\largestepprobability(\point):=1-\int\proposalkernel_\largestepprobability(\point,\dd{\otherpoint})\acceptancefunction_\largestepprobability(\point,\otherpoint)=\largestepprobability\rejectionprobability_1(\point)+(1-\largestepprobability)\rejectionprobability_0
        \end{equation} for $\point\in\measurablespace$. By definition, the generator of $(\markovchain^\largestepprobability_\discretetimepoint)_{\discretetimepoint\in\discretetimedomain}$ is \begin{equation}
            \localgenerator_\largestepprobability:=\kernel_\largestepprobability-\operatorname{id}_{\measurablesystem_b}
        \end{equation} and the generator of $(\concatenatedprocess_\timepoint)_{\timepoint\ge0}$ is \begin{equation}
            \concatenatedgenerator=\localgenerator_0+\killingrate\globalgenerator.
        \end{equation} Given that, we obtain \begin{equation}
            \begin{split}
                &\dirichletform(\integrand)-\dirichletform_\largestepprobability(\integrand)=\largestepprobability\dirichletform_0(\integrand)+\expectedlifetime_1\mathcal G(\integrand)-\largestepprobability\dirichletform_1(\integrand)\\&\;\;\;\;\ge \expectedlifetime_1\mathcal G(\integrand)-\largestepprobability\dirichletform_1(\integrand)\\&\;\;\;\;=\expectedlifetime_1\langle\integrand,\integrand-\regenerationdistribution\integrand\rangle_{L^2(\regenerationdistribution)}\\&\;\;\;\;\;\;\;\;\;\;\;\;-\frac\largestepprobability\int\targetdistribution(\dd{\otherpoint})\int\regenerationdistribution(\dd{\otherpoint})\acceptancefunction_\largestepprobability(\point,\otherpoint)\left|\integrand(\point)-\integrand(\otherpoint)\right|^2\\&\;\;\;\;\ge\expectedlifetime_1\langle\integrand,\integrand-\regenerationdistribution\integrand\rangle_{L^2(\regenerationdistribution)}\\&\;\;\;\;\;\;\;\;\;\;\;\;-\frac\largestepprobability{2\targetdensity_\referencemeasure}\int\referencemeasure(\dd{\point})\int\regenerationdistribution(\dd{\otherpoint})\left|\integrand(\point)-\integrand(\otherpoint)\right|^2\\&\;\;\;\;=\left(\expectedlifetime_1-\frac\largestepprobability{2\targetdensity_\referencemeasure}\right)\int\regenerationdistribution(\dd{\point})\int\regenerationdistribution(\dd{\otherpoint})\left|\integrand(\point)-\integrand(\otherpoint)\right|^2\\&\;\;\;\;\;\;\;\;\;\;\;\;+|\regenerationdistribution\integrand|^2-\regenerationdistribution|\integrand|^2
            \end{split}
        \end{equation} for all $\integrand\in\measurablesystem_b$, where \begin{equation}
            \mathcal G(\integrand):=-\langle\integrand,\globalgenerator\integrand\rangle_{L^2(\regenerationdistribution)},
        \end{equation} $\expectedlifetime_1$ is the constant in the definition of the killing rate $\killingrate$, and $\referencemeasure$ and $\targetdensity_\referencemeasure$ are defined as in the main paper. Nothing that $|\regenerationdistribution\integrand|^2-\regenerationdistribution|\integrand|^2$ is the variance of $\integrand$ with respect to the probability measure $\regenerationdistribution$ (hence positive for nontrivial $\integrand$), the claim is immediate from this.
    \end{proof}
\end{theorem}

\newpage
\onecolumn

\section{Additional results and error tables}\label{sec:additional-results}

\begin{figure*}
    \input{figures/kitchen}
    \captionsetup{skip = 0pt}
    \caption{Equal rendering time comparison (60\,s) of Metropolis (left), Metropolis Restore (middle), and \gls{erpt} (right) for the \textsc{Country Kitchen} scene provided by \citep{bitterli2016resources}.}
    \label{fig:kitchen}
\end{figure*}

\begin{table}[H]
	\centering
    \begin{tabular}{l r@{\,/\,}r r@{\,/\,}r r@{\,/\,}r r@{\,/\,}r r@{\,/\,}r r@{\,/\,}r}
        \toprule
        & \multicolumn{2}{c}{$L^1$-error} & \multicolumn{2}{c}{$L^2$-error} & \multicolumn{2}{c}{\gls{mrse}} & \multicolumn{2}{c}{\gls{rmse}} & \multicolumn{2}{c}{\gls{mape}} & \multicolumn{2}{c}{Variance} \\
        & \multicolumn{2}{c}{\scriptsize equal time / spp} 
        & \multicolumn{2}{c}{\scriptsize equal time / spp} 
        & \multicolumn{2}{c}{\scriptsize equal time / spp} 
        & \multicolumn{2}{c}{\scriptsize equal time / spp} 
        & \multicolumn{2}{c}{\scriptsize equal time / spp} 
        & \multicolumn{2}{c}{\scriptsize equal time / spp}
        \\
        \cmidrule(lr){2-3} \cmidrule(lr){4-5} \cmidrule(lr){6-7} \cmidrule(lr){8-9} \cmidrule(lr){10-11} \cmidrule(lr){12-13}
        Metropolis & 0.0112 & 0.0214 & 0.0022 & 0.0292 & 0.2934 & 2.0138 & 0.0197 & 0.2798 & 4.7997 & 16.6797 & 0.0014 & 0.0283 \\
        Metropolis Restore (Ours) & \textbf{0.0007} & 0.0030 & 0.0002 & 0.0091 & 0.0101 & 0.2674 & 0.0017 & 0.0980 & \textbf{2.9912} & 14.2227 & 0.0002 & 0.0091 \\
        \gls{mala} & 0.0113 & 0.0154 & 0.0025 & 0.0085 & 0.0834 & 0.1679 & 0.0221 & 0.0833 & 4.7843 & 12.1290 & 0.0014 & 0.0078 \\
        \gls{mala} Restore (Ours) & 0.0008 & \textbf{0.0023} & \textbf{0.0001} & \textbf{0.0019} & \textbf{0.0049} & \textbf{0.0356} & \textbf{0.0013} & \textbf{0.0158} & 3.4927 & \textbf{10.2949} & \textbf{0.0001} & \textbf{0.0018} \\
        \gls{hmc} & 0.0137 & 0.0245 & 0.0047 & 0.0131 & 0.4145 & 0.9282 & 0.0441 & 0.1163 & 6.6072 & 16.7781 & 0.0039 & 0.0120 \\
        \gls{hmc} Restore (Ours) & 0.0011 & 0.0046 & 0.0002 & 0.0025 & 0.0056 & 0.0608 & 0.0019 & 0.0229 & 5.0556 & 14.4194 & 0.0002 & 0.0024 \\
        \gls{erpt} & 0.0104 & 0.0149 & 0.0029 & 0.2257 & 0.0437 & 4.8108 & 0.0260 & 1.2248 & 5.5895 & 18.8111 & 0.0030 & 0.0332 \\
        \bottomrule
    \end{tabular}
	\caption{Equal rendering time (1000\,s) and equal \gls{spp} count (1000) comparison, separated by "/", of Metropolis, Metropolis Restore, \gls{mala}, \gls{mala} Restore, \gls{hmc}, \gls{hmc} Restore, and \gls{erpt} for the \textsc{Swimming Pool} scene depicted in Figure 1 of the main paper.}
	\label{tab:pool}
\end{table}

\begin{table}[H]
	\centering
    \begin{tabular}{l r@{\,/\,}r r@{\,/\,}r r@{\,/\,}r r@{\,/\,}r r@{\,/\,}r r@{\,/\,}r}
        \toprule
        & \multicolumn{2}{c}{$L^1$-error} & \multicolumn{2}{c}{$L^2$-error} & \multicolumn{2}{c}{\gls{mrse}} & \multicolumn{2}{c}{\gls{rmse}} & \multicolumn{2}{c}{\gls{mape}} & \multicolumn{2}{c}{Variance} \\
        & \multicolumn{2}{c}{\scriptsize equal time / spp} 
        & \multicolumn{2}{c}{\scriptsize equal time / spp} 
        & \multicolumn{2}{c}{\scriptsize equal time / spp} 
        & \multicolumn{2}{c}{\scriptsize equal time / spp} 
        & \multicolumn{2}{c}{\scriptsize equal time / spp} 
        & \multicolumn{2}{c}{\scriptsize equal time / spp}
        \\
        \cmidrule(lr){2-3} \cmidrule(lr){4-5} \cmidrule(lr){6-7} \cmidrule(lr){8-9} \cmidrule(lr){10-11} \cmidrule(lr){12-13}
        Metropolis & 0.0096 & 0.0314 & 0.0007 & 0.0037 & 0.0044 & 0.0773 & 0.0009 & 0.0050 & 4.1148 & 14.9918 & 0.0002 & 0.0032 \\
        Metropolis Restore (Ours) & 0.0012 & 0.0036 & 0.0002 & 0.0005 & 0.0005 & 0.0074 & 0.0003 & 0.0007 & 3.6815 & 13.3062 & \textbf{0.0001} & 0.0004 \\
        \gls{mala} & 0.0082 & 0.0230 & 0.0007 & 0.0025 & 0.0027 & 0.0368 & 0.0009 & 0.0034 & 3.2982 & 10.6530 & 0.0001 & 0.0019 \\
        \gls{mala} Restore (Ours) & \textbf{0.0011} & \textbf{0.0025} & \textbf{0.0002} & \textbf{0.0003} & \textbf{0.0004} & \textbf{0.0036} & \textbf{0.0002} & \textbf{0.0004} & \textbf{3.0106} & \textbf{9.5083} & 0.0001 & \textbf{0.0002} \\
        \gls{hmc} & 0.0116 & 0.0302 & 0.0008 & 0.0033 & 0.0065 & 0.0612 & 0.0010 & 0.0046 & 5.0588 & 14.2838 & 0.0002 & 0.0028 \\
        \gls{hmc} Restore (Ours) & 0.0015 & 0.0035 & 0.0002 & 0.0006 & 0.0007 & 0.0058 & 0.0003 & 0.0008 & 4.8540 & 12.8438 & 0.0001 & 0.0004 \\
        \gls{erpt} & 0.0042 & 0.0082 & 0.0006 & 0.0036 & 0.0022 & 0.0204 & 0.0008 & 0.0049 & 11.5261 & 22.2709 & 0.0016 & 0.0046 \\
        \bottomrule
    \end{tabular}
	\caption{Equal rendering time (1000\,s) and equal \gls{spp} count (1000) comparison, separated by "/", of Metropolis, Metropolis Restore, \gls{mala}, \gls{mala} Restore, \gls{hmc}, \gls{hmc} Restore, and \gls{erpt} for the \textsc{Veach, ajar} scene depicted in Figure 3 of the main paper.}
	\label{tab:torus}
\end{table}

\begin{table}[H]
	\centering
    \begin{tabular}{l r@{\,/\,}r r@{\,/\,}r r@{\,/\,}r r@{\,/\,}r r@{\,/\,}r r@{\,/\,}r}
        \toprule
        & \multicolumn{2}{c}{$L^1$-error} & \multicolumn{2}{c}{$L^2$-error} & \multicolumn{2}{c}{\gls{mrse}} & \multicolumn{2}{c}{\gls{rmse}} & \multicolumn{2}{c}{\gls{mape}} & \multicolumn{2}{c}{Variance} \\
        & \multicolumn{2}{c}{\scriptsize equal time / spp} 
        & \multicolumn{2}{c}{\scriptsize equal time / spp} 
        & \multicolumn{2}{c}{\scriptsize equal time / spp} 
        & \multicolumn{2}{c}{\scriptsize equal time / spp} 
        & \multicolumn{2}{c}{\scriptsize equal time / spp} 
        & \multicolumn{2}{c}{\scriptsize equal time / spp}
        \\
        \cmidrule(lr){2-3} \cmidrule(lr){4-5} \cmidrule(lr){6-7} \cmidrule(lr){8-9} \cmidrule(lr){10-11} \cmidrule(lr){12-13}
        Metropolis & 0.0071 & 0.0165 & 0.0003 & 0.0027 & 0.0080 & 0.1365 & 0.0056 & 0.0431 & 4.4235 & 11.0185 & 0.0002 & 0.0024 \\
        Metropolis Restore (Ours) & 0.0013 & 0.0020 & 0.0001 & 0.0003 & \textbf{0.0011} & 0.0083 & 0.0012 & 0.0042 & 6.4766 & 10.5329 & \textbf{0.0001} & 0.0002 \\
        \gls{mala} & 0.0070 & 0.0138 & 0.0003 & 0.0016 & 0.0059 & 0.0499 & 0.0052 & 0.0258 & \textbf{4.2916} & 9.2475 & 0.0002 & 0.0014 \\
        \gls{mala} Restore (Ours) & \textbf{0.0011} & \textbf{0.0017} & 0.0001 & \textbf{0.0002} & 0.0012 & \textbf{0.0079} & 0.0011 & \textbf{0.0034} & 5.5375 & \textbf{8.8199} & 0.0001 & 0.0002 \\
        \gls{hmc} & 0.0071 & 0.0139 & 0.0004 & 0.0015 & 0.0121 & 0.0693 & 0.0064 & 0.0241 & 4.7984 & 9.4230 & 0.0002 & 0.0012 \\
        \gls{hmc} Restore (Ours) & 0.0012 & 0.0020 & \textbf{0.0001} & 0.0003 & 0.0016 & 0.0120 & \textbf{0.0008} & 0.0046 & 6.6775 & 9.3272 & 0.0001 & \textbf{0.0002} \\
        \gls{erpt} & 0.0073 & 0.0102 & 0.0008 & 0.0170 & 0.0116 & 0.3746 & 0.0137 & 0.2717 & 8.8943 & 14.0592 & 0.0004 & 0.0166 \\
        \bottomrule
    \end{tabular}
	\caption{Equal rendering time (1000\,s) and equal \gls{spp} count (1000) comparison, separated by "/", of Metropolis, Metropolis Restore, \gls{mala}, \gls{mala} Restore, \gls{hmc}, \gls{hmc} Restore, and \gls{erpt} for the \textsc{Torus} scene depicted in Figure 5 of the main paper.}
	\label{tab:torus}
\end{table}

\begin{table}[H]
	\centering
    \begin{tabular}{l r@{\,/\,}r r@{\,/\,}r r@{\,/\,}r r@{\,/\,}r r@{\,/\,}r r@{\,/\,}r}
        \toprule
        & \multicolumn{2}{c}{$L^1$-error} & \multicolumn{2}{c}{$L^2$-error} & \multicolumn{2}{c}{\gls{mrse}} & \multicolumn{2}{c}{\gls{rmse}} & \multicolumn{2}{c}{\gls{mape}} & \multicolumn{2}{c}{Variance} \\
        & \multicolumn{2}{c}{\scriptsize equal time / spp} 
        & \multicolumn{2}{c}{\scriptsize equal time / spp} 
        & \multicolumn{2}{c}{\scriptsize equal time / spp} 
        & \multicolumn{2}{c}{\scriptsize equal time / spp} 
        & \multicolumn{2}{c}{\scriptsize equal time / spp} 
        & \multicolumn{2}{c}{\scriptsize equal time / spp}
        \\
        \cmidrule(lr){2-3} \cmidrule(lr){4-5} \cmidrule(lr){6-7} \cmidrule(lr){8-9} \cmidrule(lr){10-11} \cmidrule(lr){12-13}
        Metropolis & 0.7559 & 0.1862 & 5.7071 & 0.8791 & 6.4681 & 2.8307 & 0.1047 & 0.0161 & 8.3683 & 5.3221 & 0.6764 & 0.8624 \\
        Metropolis Restore (Ours) & \textbf{0.0788} & \textbf{0.1324} & \textbf{0.0745} & \textbf{0.1949} & \textbf{1.1649} & \textbf{1.7006} & \textbf{0.0014} & \textbf{0.0036} & \textbf{3.9425} & \textbf{3.1927} & \textbf{0.0573} & \textbf{0.1829} \\
        \gls{erpt} & 0.8317 & 2.7287 & 0.5215 & 5.7336 & 5.9559 & 1.7238 & 0.0096 & 0.1051 & 9.6347 & 4.9365 & 0.1624 & 5.0586 \\
        \bottomrule
    \end{tabular}
	\caption{Equal rendering time (1000\,s) and equal \gls{spp} count (1000) comparison, separated by "/", of Metropolis, Metropolis Restore, \gls{mala}, \gls{mala} Restore, \gls{hmc}, \gls{hmc} Restore, and \gls{erpt} for the \textsc{Contemporary Bathroom} scene depicted in Figure 6 of the main paper.}
	\label{tab:torus}
\end{table}

\begin{table}[H]
	\centering
    \begin{tabular}{l r@{\,/\,}r r@{\,/\,}r r@{\,/\,}r r@{\,/\,}r r@{\,/\,}r r@{\,/\,}r}
        \toprule
        & \multicolumn{2}{c}{$L^1$-error} & \multicolumn{2}{c}{$L^2$-error} & \multicolumn{2}{c}{\gls{mrse}} & \multicolumn{2}{c}{\gls{rmse}} & \multicolumn{2}{c}{\gls{mape}} & \multicolumn{2}{c}{Variance} \\
        & \multicolumn{2}{c}{\scriptsize equal time / spp} 
        & \multicolumn{2}{c}{\scriptsize equal time / spp} 
        & \multicolumn{2}{c}{\scriptsize equal time / spp} 
        & \multicolumn{2}{c}{\scriptsize equal time / spp} 
        & \multicolumn{2}{c}{\scriptsize equal time / spp} 
        & \multicolumn{2}{c}{\scriptsize equal time / spp}
        \\
        \cmidrule(lr){2-3} \cmidrule(lr){4-5} \cmidrule(lr){6-7} \cmidrule(lr){8-9} \cmidrule(lr){10-11} \cmidrule(lr){12-13}
        Metropolis & 0.1371 & 0.0332 & 0.3204 & 0.0215 & 1.9822 & 0.8813 & 0.5793 & 0.0397 & 5.4077 & 12.4250 & 0.2729 & 0.0209 \\
        Metropolis Restore (Ours) & \textbf{0.0107} & \textbf{0.0243} & \textbf{0.0021} & \textbf{0.0166} & \textbf{0.0985} & \textbf{0.7711} & \textbf{0.0039} & \textbf{0.0301} & \textbf{4.5303} & \textbf{10.0534} & \textbf{0.0016} & \textbf{0.0157} \\
        \gls{erpt} & 0.1612 & 0.2001 & 0.0630 & 0.1277 & 0.2332 & 1.2607 & 0.1753 & 0.2431 & 4.9591 & 18.0779 & 0.0157 & 0.0525 \\
        \bottomrule
    \end{tabular}
	\caption{Equal rendering time (1000\,s) and equal \gls{spp} count (1000) comparison, separated by "/", of Metropolis, Metropolis Restore, \gls{mala}, \gls{mala} Restore, \gls{hmc}, \gls{hmc} Restore, and \gls{erpt} for the \textsc{Glass of Water} scene depicted in Figure 8 of the main paper.}
	\label{tab:torus}
\end{table}

\begin{table}[H]
	\centering
    \begin{tabular}{l r@{\,/\,}r r@{\,/\,}r r@{\,/\,}r r@{\,/\,}r r@{\,/\,}r r@{\,/\,}r}
        \toprule
        & \multicolumn{2}{c}{MAE} & \multicolumn{2}{c}{MSE} & \multicolumn{2}{c}{\gls{mrse}} & \multicolumn{2}{c}{\gls{rmse}} & \multicolumn{2}{c}{\gls{mape}} & \multicolumn{2}{c}{Empirical Variance} \\
        & \multicolumn{2}{c}{\scriptsize equal time / sample count} 
        & \multicolumn{2}{c}{\scriptsize equal time / sample count} 
        & \multicolumn{2}{c}{\scriptsize equal time / sample count} 
        & \multicolumn{2}{c}{\scriptsize equal time / sample count} 
        & \multicolumn{2}{c}{\scriptsize equal time / sample count} 
        & \multicolumn{2}{c}{\scriptsize equal time / sample count}
        \\
        \cmidrule(lr){2-3} \cmidrule(lr){4-5} \cmidrule(lr){6-7} \cmidrule(lr){8-9} \cmidrule(lr){10-11} \cmidrule(lr){12-13}
        Baseline & 0.2520 & 0.0719 & 0.4290 & 0.0382 & 0.0686 & 0.4145 & 0.0290 & 0.0026 & 8.3453 & 14.3491 & 0.0249 & 0.0378 \\
        Ours & \textbf{0.0279} & \textbf{0.0539} & \textbf{0.0037} & \textbf{0.0157} & \textbf{0.0285} & \textbf{0.1405} & \textbf{0.0002} & \textbf{0.0011} & \textbf{7.6564} & \textbf{9.5531} & \textbf{0.0008} & \textbf{0.0126} \\
        \gls{erpt} & 0.0477 & 0.1290 & 0.0103 & 0.1267 & 0.0395 & 0.3354 & 0.0032 & 0.0086 & 7.8014 & 13.6678 & 0.0016 & 0.0748 \\
        \bottomrule
    \end{tabular}
	\caption{Equal rendering time (1000\,s) and equal \gls{spp} count (1000) comparison, separated by "/", of Metropolis, Metropolis Restore, \gls{mala}, \gls{mala} Restore, \gls{hmc}, \gls{hmc} Restore, and \gls{erpt} for the \textsc{Country Kitchen} scene depicted in \autoref{fig:kitchen}.}
	\label{tab:torus}
\end{table}

\fi

\end{document}